\documentclass[lettersize,journal]{IEEEtran}
\ifCLASSINFOpdf
\else
   \usepackage{graphicx}
   \graphicspath{{../eps/}}
   \DeclareGraphicsExtensions{.eps}
\fi

\usepackage{amsmath}
\usepackage{amsfonts,amssymb}
\usepackage{amssymb}
\usepackage{wrapfig}
\usepackage{psfrag}
\usepackage{epstopdf}
\usepackage{cite}
\usepackage{graphicx}
\usepackage{float} 
\usepackage{subfigure}
\usepackage{threeparttable}
\usepackage{cases}
\usepackage{subeqnarray}
\usepackage{color}
\usepackage{underscore}
\usepackage{verbatim}
\usepackage{bm}
\usepackage{stfloats}
\usepackage{ragged2e} 
\usepackage{diagbox}
\usepackage{multirow}     
\usepackage{longtable}
\usepackage{url}
\usepackage[normalem]{ulem}
\usepackage{tabularray}
\newtheorem{theorem}{Theorem}

\newtheorem{algorithm}{Algorithm}

\usepackage{algorithm}
\usepackage{algorithmicx}
\usepackage{algpseudocode}

\begin{document}

\title{AI-based Environment-Aware XL-MIMO Channel Estimation with Location-Specific Prior Knowledge Enabled by CKM}

\author{Yuelong~Qiu, Di~Wu, Yong~Zeng,~\IEEEmembership{Fellow,~IEEE}, Yanqun~Tang, 

Nan~Cheng,~\IEEEmembership{Senior Member,~IEEE}, and Chenhao Qi,~\IEEEmembership{Senior Member,~IEEE}
        % <-this % stops a space
\thanks{Y. Qiu, D. Wu, Y. Zeng, and C. Qi are with the National Mobile Communications Research Laboratory, Southeast University, Nanjing 210096, China (e-mail: {yl_qiu, studywudi, yong_zeng, qch}@seu.edu.cn). 

Y. Zeng is also with the Purple Mountain Laboratories, Nanjing 211111, China (corresponding author: Y. Zeng)

Y. Tang is with School of Electronics and Communication Engineering, Sun Yat-Sen University, Guangzhou, 510006, China (email: tangyq8@mail.sysu.edu.cn).

N. Cheng is with the School of Telecommunications Engineering, Xidian University, Xi'an 710071, China (e-mail: nancheng@xidian.edu.cn).}% <-this % stops a space
}

\maketitle

\begin{abstract}
Accurate and efficient acquisition of wireless channel state information (CSI) is crucial to enhance the communication performance of wireless systems. However, with the continuous densification of wireless links, increased channel dimensions, and the use of higher-frequency bands, channel estimation in the sixth generation (6G) and beyond wireless networks faces new challenges, such as insufficient orthogonal pilot sequences, inadequate signal-to-noise ratio (SNR) for channel training, and more sophisticated channel statistical distributions in complex environment. These challenges pose significant difficulties for classical channel estimation algorithms like least squares (LS) and maximum a posteriori (MAP). To address this problem, we propose a novel environment-aware channel estimation framework with location-specific prior channel distribution enabled by the new concept of channel knowledge map (CKM). To this end, we propose a new type of CKM called channel score function map (CSFM), which learns the channel probability density function (PDF) using artificial intelligence (AI) techniques. To fully exploit the prior information in CSFM, we propose a plug-and-play (PnP) based algorithm to decouple the regularized MAP channel estimation problem, thereby reducing the complexity of the optimization process. Besides, we employ Tweedie’s formula to establish a connection between the channel score function, defined as the logarithmic gradient of the channel PDF, and the channel denoiser. This allows the use of the high-precision, environment-aware channel denoiser from the CSFM to approximate the channel score function, thus enabling efficient processing of the decoupled channel statistical components. Simulation results show that the proposed CSFM-PnP based channel estimation technique significantly outperforms the conventional techniques in the aforementioned challenging scenarios.
\end{abstract}
\begin{IEEEkeywords}  
Large-dimensional channel estimation, environment-aware channel estimation, channel knowledge map, channel score function map.
\end{IEEEkeywords}

% \fi
\IEEEpeerreviewmaketitle

\section{Introduction}
Accurate and efficient acquisition of wireless channel state information (CSI) is essential for the design of the sixth generation (6G) and beyond mobile communication networks, as it governs fundamental communication performance and impacts the achievability of key metrics like ultra-high data rates, ultra-low latency, and ultra-high reliability \cite{CKMTutorial}. However, with the continuous densification of wireless links, increased channel dimensions, and the use of higher-frequency bands, channel estimation will face new challenges. Firstly, the number of orthogonal pilot sequences will become insufficient due to the significant growth in antenna dimensions and continuous densification of wireless links  \cite{LimitedPilot, LimitedPilotUEs}. Secondly, the signal-to-noise ratio (SNR) for channel training may become inadequate with the use of millimeter-wave (mmWave) and terahertz (THz) signals which suffers from severe path loss, significantly reducing the SNR at the receiver. Thirdly, channel statistical distributions become more sophisticated in complex environments and typical assumptions like Rayleigh, Rician, and Nakagami distributions may not be accurate enough, as their properties are highly dependent on the wireless \mbox{environment \cite{DTC}}. In more general cases, it is very difficult to obtain a closed-form expression for the probability density function (PDF) of channel. Lastly, in extreme scenarios, we may even wan to infer the channel without using any pilot signal to achieve the so-called channel training-free communication, say in highly dynamic scenarios where the channel coherence time may be extremely short, insufficient to support real-time channel estimation based on pilot signals \cite{WithoutPilot}.

Channel estimation for extremely large-scale multiple-input multiple-output (XL-MIMO) systems under the four challenging scenarios outlined above presents significant \mbox{difficulties \cite{XLMIMOTutorial}}. In practical XL-MIMO wirless communication systems, both far-field and near-field regions often exist simultaneously. Consequently, the study of hybrid-field channel estimation is more universal, and it is primarily categorized into two categories. The first category separates the far-field and near-field components using channel statistical knowledge or prior information, then applies classical algorithms designed for the far-field and near-field channels, \mbox{respectively \cite{XLMIMOFNH, XLMIMOH}}. However, this separation would be greatly impacted by the aforementioned four challenging conditions. The second category consists of classical channel estimation algorithms, including least \mbox{squares (LS) \cite{LS, LS2}}, maximum likelihood (ML) \cite{ML}, minimum mean square error (MMSE) \cite{SSP, MMSE}, linear minimum mean square error (LMMSE) \cite{SSP, LMMSE}, and maximum a posteriori (MAP) \cite{SSP, MAP}. However, in challenging scenarios with insufficient orthogonal pilot sequences, inadequate training SNR, or no pilot signals, the quality of effective channel information in real-time observations is extremely poor, severely degrading the performance of the five algorithms. Additionally, complex channel prior distributions in complicated environments deteriorate the performance of the LMMSE algorithm. Meanwhile, the MMSE and MAP algorithms face difficulties in obtaining complex channel prior information and have high computational complexity.

Compared with LS, ML, LMMSE, and MMSE, the MAP algorithm offers distinct advantages in the four aforementioned challenging scenarios. Specifically, the MAP algorithm leverages the PDF of the channel, resulting in better performance than LS and ML algorithms. Moreover, by utilizing the complete PDF of the channel, MAP outperforms LMMSE, which relies solely on first-order (mean) and second-order (covariance) channel statistics information. Additionally, compared with MMSE, MAP exhibits lower computational complexity and avoids the ``regression to the mean'' issue, thereby preventing the loss of critical details and ensuring that the inferred solution remains within the expected solution space \cite{InverseProblem}.

It is expected to address the difficulties of MAP algorithm through emerging technologies such as Plug-and-play \mbox{(PnP) \cite{PnP}} and artificial intelligence (AI). When the noise term in the pilot observation signal is modeled as the commonly used Gaussian distribution, the MAP channel estimation problem can be maximally simplified using Bayes rule \cite{SSP}. The simplified optimization problem includes both the pilot observation term and the channel prior distribution term. However, in complex environments where the PDF of the channel is intricate, directly solving the aforementioned optimization problem is impractical. Therefore, by employing the PnP concept, the pilot observation term and the channel distribution term in the MAP optimization problem are separated into two subproblems, which are processed independently and iterated until convergence to yield the final channel estimation \cite{PnP}. Specifically, when the PDF of channel can be expressed in a closed form and is differentiable, the channel distribution term can be solved using various gradient descent \mbox{algorithms \cite{Convexoptimization}}. When the channel PDF is difficult to express in closed form, the channel prior distribution term can be effectively approximated by leveraging the channel score function \cite{ScoreFunction}, Tweedie's formula \cite{Tweedie}, and channel denoiser, inspired by advances in the field of image processing. The channel score function is defined as the gradient of the log channel distribution. Tweedie's formula links the score function to the channel denoiser, implying that the log gradient of any channel PDF can be approximated with high precision using a high-quality channel denoiser. And the authors in \cite{AIScore} and \cite{DeepPnP} have designed channel denoisers using AI techniques based on historical channel datasets collected across the entire wireless communication environment. However, these methods fail to account for the strong coupling between channel characteristics and the wireless environment, which may lead to inaccurate approximations of channel score functions at different locations and consequently degrade channel estimation accuracy.

The CKM proposed in \cite{CKMMagazine} serves as a crucial enabler for environment-aware communication and sensing, offering an innovative solution for the efficient acquisition of the environment-aware channel score function. By integrating historical channel knowledge data within a region, CKM constructs a channel knowledge database that reflects the intrinsic characteristics of wireless channels. It can directly obtain environmental prior information based on the physical or virtual locations of the mobile terminals, thereby avoiding redundant sensing of static environmental components or repeated channel estimation. In recent years, CKM has been widely studied to enhance communication and sensing. For CKM-assisted communication, existing research primarily focuses on beam alignment, channel estimation, and resource allocation. Training-free beam alignment under a pure analog architecture was achieved in \cite{CKMBF}. Subsequently, this approach was extended to hybrid analog-digital and intelligent reflecting surface (IRS) scenarios in \cite{CKMBFH} and \cite{CKMIRS}, respectively. Meanwhile, channel estimation for dynamic scenario \cite{CKMDynamic} and multi-user scenario \cite{CCM} was implemented. Additionally, efficient communication resource allocation was realized by utilizing prior knowledge in CKM \cite{CKMRA}. For CKM-assisted sensing, fingerprinting based on joint angle, delay, and other multi-dimensional channel knowledge was realized in \cite{CKMFP}. The concept of clutter angular map (CLAM) is proposed to effectively improve sensing performance under low SNR \mbox{conditions \cite{CLAM}}. CKM significantly diverges from traditional methods of acquiring CSI by explicitly highlighting the strong correlation between wireless channel characteristics and environmental features at specific locations, thereby reducing the uncertainty associated with channel prior distribution information.

Motivated by the above discussions, this paper proposes a novel AI-based environment-aware XL-MIMO channel estimation framework with location-specific prior channel distribution enabled by CKM. The main contributions of this paper are summarized as follows: 

$\bullet$ For the problem of high-quality large-dimensional wireless channel estimation under challenging scenarios, such as insufficient orthogonal pilot sequences, inadequate SNR for channel training, and sophisticated channel statistical distributions in complex environments, this paper proposes a regularized MAP estimation framework to achieve low-cost, high-precision channel estimation.

$\bullet$ To address the challenge of solving the regularized MAP optimization problem under complex channel distributions, this paper employs the PnP method to decompose the original problem into two sub-optimization problems. The first sub-problem, which includes the pilot observation term, is a quadratic programming problem that can be solved directly. The second sub-problem, which involves the complex channel prior distribution term, is efficiently solved via the steepest descent (SD) method by using the channel score function. The channel score function is approximated by the environment-aware channel denoiser provided by a new type of CKM called channel score function map (CSFM). Finally, the channel estimation is achieved by iteratively solving these two sub-problems until convergence.

$\bullet$ Simulation results demostrate that the proposed environment-aware channel estimation algorithm exhibits a notable superiority over traditional baseline channel estimation algorithms such as LS, ML, and LMMSE under the aforementioned challenging scenarios.

The remainder of this paper is organized as follows. Section II introduces the system model. Section III presents the classic channel estimation methods and their limitations. Section IV and Section V present the proposed channel estimation algorithm and corresponding CSFM construction method, respectively. Section VI shows the simulation results and Section VII draws the conclusion.

\textbf{Notations:} Boldface lower- and upper-case letters denote a column vector and a matrix, respectively. Non-bold letter $a$ and $A$ are scalars. $\mathbb{E}$ denotes the statistical expectation. Caligraphic letter $\mathcal{A}$ is the mapping. $\left\|\mathbf{a}\right\|$ is the 2-norm of column vector $\mathbf{a}$ and $\left\|\mathbf{A}\right\|_F$ is the Frobenius norm of matrix $\mathbf{A}$. $\mathbf{A}^T$, $\mathbf{A}^{H}$, $\mathbf{A}^{-1}$, and $\mathbf{A}^\dagger$ are the transpose, Hermitian, inverse, and pseudo inverse of $\mathbf{A}$, respectively. $\nabla_{\mathbf{A}}\left(\cdot\right)|_{\mathbf{A}=\mathbf{B}}$ represents the partial derivative of the parameter $\mathbf{A}$ at value $\mathbf{B}$. $tr\left(\mathbf{A}\right)$ denotes the trace of the square matrix $\mathbf{A}$.

\section{System Model}
\begin{figure}[htbp]
\centering
\includegraphics[scale=0.085]{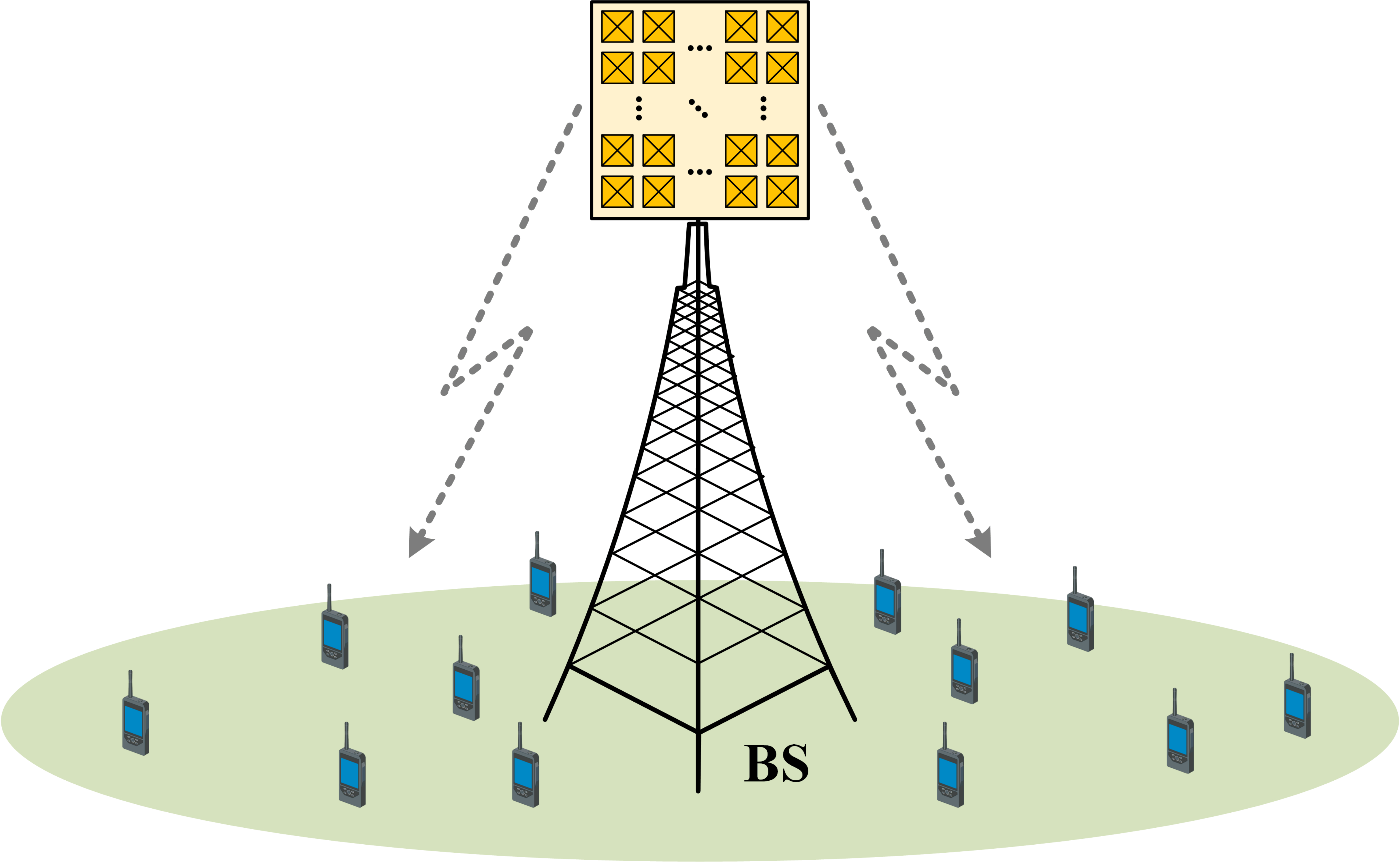}
\caption{XL-MIMO DL wireless communication system.}
\label{figMIMODL}
\end{figure}
Consider a XL-MIMO downlink (DL) wireless communication system depicted in Fig. \ref{figMIMODL}, where the base station (BS) equipped with $M\gg 1$ antennas transmits wireless signals to $K$ single-antenna user equippments (UEs). For classic DL channel estimation, the BS sends orthogonal pilot sequences to the UEs over $\tau$ symbol durations, and each UE performs channel estimation independently \cite{LimitedPilot, DLChannelEstimation}. The pilot signal received by UE $k$ at symbol duration $n$ can be expressed as
\begin{equation}
y_k\left[n\right]=\sqrt{\rho\xi_k}\mathbf{h}_k^{T}\mathbf{x}\left[n\right]+z_k\left[n\right],n=1,2,...,\tau, \label{eq1}
\end{equation}
where $y_k\left[n\right]$ is the received pilot signal and $\mathbf{x}\left[n\right]=\left[x_{1}\left[n\right],x_{2}\left[n\right],...,x_{M}\left[n\right]\right]^T\in \mathbb{C}^{M\times 1}$ is the transmitted pilot signals. $\xi_k$ is the large-scale channel power gain of the channel between the BS and UE $k$. $\mathbf{h}_k \in \mathbb{C}^{M\times 1}$ is the normalized small-scale  channel vector with $\left\|\mathbf{h}_k\right\|=1$. Besides, $\rho$ is the transmitted pilot power and $z_k\left[n\right]$ is zero-mean circularly symmetric complex Gaussian (CSCG) noise with variance $\sigma_{k}^{2}$.

By concatenating the transmitted and received signals across these $\tau$ symbol durations, we have
\begin{equation}
\mathbf{y}_k=\sqrt{\rho\xi_k}\mathbf{X}\mathbf{h}_k+\mathbf{z}_k, \label{eq2}
\end{equation}
where $\mathbf{y}_k=\left[y_k\left[1\right], y_k\left[2\right],..., y_k\left[\tau\right]\right]^{T}\in \mathbb{C}^{\tau\times 1}$, $\mathbf{X}=$ $\left[\mathbf{x}\left[1\right], \mathbf{x}\left[2\right],..., \mathbf{x}\left[\tau\right]\right]^{T}\in \mathbb{C}^{\tau\times M}$, and $\mathbf{z}_k=[z_k\left[1\right], z_k\left[2\right],...,$ $ z_k\left[\tau\right]]^{T}\in \mathbb{C}^{\tau\times 1}$. The transmitted pilot signals satisfy the energy constraint $\left\|\sqrt{\rho}\mathbf{X}\right\|^2_F=\rho\tau$. Besides, according to equation \eqref{eq2}, the received expected SNR of UE $k$ can be defined as
\begin{equation}
SNR_k=\frac{\mathbb{E}\left\{\left|\left|\sqrt{\rho\xi_k}\mathbf{X}\mathbf{h}_k\right|\right|^2\right\}}{\mathbb{E}\left\{\left|\left|\mathbf{z}_k\right|\right|^2\right\}}=\frac{\rho\xi_k tr\left(\mathbf{X}\mathbb{E}\left\{\mathbf{h}_k\mathbf{h}_k^H\right\}\mathbf{X}^H\right)}{\tau\sigma_k^2}. \label{eq3}
\end{equation}

Channel estimation based on equation \eqref{eq2} is a classic linear inverse problem that provides a unified framework for various scenarios:

$\bullet$ Pilot-based Channel Estimation: When $\tau>0$, the received signal $\mathbf{y}_k$ in equation \eqref{eq2} comprises both the received pilot signal $\sqrt{\rho\xi_k}\mathbf{X}\mathbf{h}_k$ and the noise $\mathbf{z}_{k}$.

$\bullet$ Channel Generation: When $\tau=0$, it signifies that no pilot signal has been transmitted from the BS to UE $k$. In this case, equation \eqref{eq2} simplifies to $\mathbf{y}_k=\mathbf{z}_{k}$. Therefore, estimating the channel based on equation \eqref{eq2} is equivalent to generating the channel from Gaussian noise, which well aligns the generative AI task \cite{GenerativeAI}.

\section{Classic Channel Estimation Methods and Their Limitations}
Since DL channel estimations for different UEs can be applied similarly \cite{LimitedPilot, DLChannelEstimation}, the subsequent analysis will omit the subscript $k$ in equation \eqref{eq2}. For channel estimation based on equation \eqref{eq2}, classic methods include LS \cite{LS, LS2}, \mbox{ML \cite{ML}}, LMMSE \cite{LMMSE}, MMSE \cite{MMSE}, and MAP \cite{MAP}. These can be categorized into two groups based on whether they use prior channel statistical distribution $P\left(\mathbf{h}\right)$ or not. Specifically, LS and ML do not rely on any prior channel distribution, while MMSE, LMMSE, and MAP need it. These methods usually work well when there are enough orthogonal pilot resources ($\tau\ge M$), high SNR, and Gaussian groundtruth channel distribution. However, in the complex and dynamic large-dimensional wireless systems of 6G, these classic algorithms encounter significant challenges, particularly in the following scenarios: insufficient orthogonal pilots, inadequate SNR, sophisticated groundtruth channel statistical distribution, and no pilot signal.

\subsection{Channel Estimation Without Channel Statistical Knowledge}
\subsubsection{LS Channel Estimation}
Based on equation \eqref{eq2}, the LS channel estimation method is
\begin{equation}
\widehat{\mathbf{h}}_{\text{LS}}=\underset{\mathbf{h}}{\mathop{\text{arg}\min }}\, \left|\left|\sqrt{\rho\xi}\mathbf{Xh}-\mathbf{y}\right|\right|^2.\label{eq4}
\end{equation}
Evidently, the LS algorithm does not utilize the channel PDF $P_h\left(\mathbf{h}\right)$, nor the likelihood function $P_{y|h}\left(\mathbf{y}|\mathbf{h}\right)$. Based on \eqref{eq4}, the expression for the LS channel estimation is \cite{LS}:
\begin{equation}
\widehat{\mathbf{h}}_{\text{LS}}=\frac{1}{\sqrt{\rho\xi}}\mathbf{X}^\dagger\mathbf{y}=\mathbf{X}^\dagger\mathbf{Xh}+\frac{1}{\sqrt{\rho\xi}}\mathbf{X}^\dagger\mathbf{z}. \label{eq5}
\end{equation}
Note that when the number of orthogonal pilot sequences is no smaller than the number of antennas at the BS, i.e., $\tau\ge M$, the pseudoinverse $\mathbf{X}^\dagger$ in \eqref{eq5} is given by $\left(\mathbf{X}^{H}\mathbf{X}\right)^{-1}\mathbf{X}^{H}$. In this case, $\widehat{\mathbf{h}}_{\text{LS}}$ is given by
\begin{equation}
\widehat{\mathbf{h}}_{\text{LS}}=\mathbf{h}+\frac{1}{\sqrt{\rho\xi}}\mathbf{X}^\dagger\mathbf{z}.\label{eq6}
\end{equation}

On the other hand, when $\tau< M$, equation \eqref{eq2} is an underdetermined function, which has an infinite number of solutions. In this scenario, by imposing the constraint of minimizing the Euclidean norm of the channel vector $\mathbf{h}$, a unique solution can still be obtained. And the pseudo-inverse in equation \eqref{eq5} is given by $\mathbf{X}^\dagger=\mathbf{X}^{H}\left(\mathbf{X}\mathbf{X}^{H}\right)^{-1}$ \cite{LSUD}.

The challenges faced by the LS channel estimation algorithm are as follows:

$\bullet$ Insufficient orthogonal pilots ($0<\tau\ll M$): In this case, $\widehat{\mathbf{h}}_{\text{LS}}$ can be obtained by equation \eqref{eq5}. However, compared to situations where the number of pilot signals is sufficient, the use of a limited number of orthogonal pilots results in rather poor channel estimation performance.

$\bullet$ Inadequate SNR: Since the received signal $\mathbf{y}$ primarily consists of noise, $\widehat{\mathbf{h}}_{\text{LS}}\approx \frac{1}{\sqrt{\rho\xi}}\mathbf{X}^\dagger\mathbf{z}$ in this case.

$\bullet$ No pilot signal ($\tau=0$): In this case, $\mathbf{X}=\mathbf{0}$. According to \eqref{eq4}, no meaningful channel estimation/generation results would be obtained.

\subsubsection{ML Channel Estimation}
For equation \eqref{eq2}, the principle of ML estimation is to identify $\mathbf{h}$ such that the likelihood function $P_{y|h}\left(\mathbf{y}|\mathbf{h}\right)$ attains its maximum value. The channel estimation result of ML algorithm can be expressed as follows
\begin{equation}
\begin{split}
\widehat{\mathbf{h}}_{\rm ML} &= \arg\max_{\mathbf{h}} P_{y|h}\left(\mathbf{y}|\mathbf{h}\right)\\
&=\arg\max_{\mathbf{h}}P_{z}\left(\mathbf{y}-\sqrt{\rho\xi}\mathbf{Xh}\right), \label{eq7}
\end{split}
\end{equation}
where $P_{z}\left(\cdot\right)$ is the PDF of the noise vector $\mathbf{z}$. Evidently, compared to the LS algorithm, the ML algorithm requires the $P_{y|h}\left(\mathbf{y}|\mathbf{h}\right)$. Specifically, $\widehat{\mathbf{h}}_{\rm ML}=\widehat{\mathbf{h}}_{\rm LS}$ when $\mathbf{z}\sim \mathcal{CN}\left(\mathbf{0},\sigma^{2}\mathbf{I}\right)$. Consequently, for the scenario considered in this paper, the ML algorithm faces the same challenges as the LS algorithm.

\subsection{Channel Estimation With Channel Statistical Knowledge}
\subsubsection{LMMSE Channel Estimation}
The LMMSE channel estimation method assumes that the channel vector $\mathbf{h}$ can be linearly represented based on the observation signal $\mathbf{y}$. The linear weighting coefficients are obtained by minimizing the mean square error (MSE) between the estimated channel $\widehat{\mathbf{h}}_{\text{LMMSE}}$ and the groundtruth channel $\mathbf{h}$. The channel estimation result of LMMSE is \cite{SSP}
\begin{equation}
\begin{split}
\widehat{\mathbf{h}}_{\text {LMMSE }}=&\overline{\mathbf{h}}+\sqrt{\rho\xi}\mathbf{C}_{\mathbf{h}}\mathbf{X}^{H}\left(\rho\xi\mathbf{X C}_{\mathbf{h}}\mathbf{X}^{H}+\sigma^{2} \mathbf{I}_{\tau}\right)^{-1}\cdot\\
&\left(\mathbf{y}-\sqrt{\rho\xi}\mathbf{X} \overline{\mathbf{h}}\right), \label{eq8}
\end{split}
\end{equation}
where $\overline{\mathbf{h}}\triangleq\mathbb{E}\left\{\mathbf{h}\right\}$ and $\mathbf{C}_{\mathbf{h}}\triangleq\mathbb{E}\left\{\left(\mathbf{h}-\overline{\mathbf{h}}\right)\left(\mathbf{h}-\overline{\mathbf{h}}\right)^{H}\right\}$ are the mean vector and covariance matrix of the channel vector $\mathbf{h}$, respectively. It should be noted that when $\mathbf{h}\sim\mathcal{CN}\left(\overline{\mathbf{h}},\mathbf{C}_{\mathbf{h}}\right)$, the LMMSE algorithm yields an optimal solution in term of minimizing the MSE \cite{SSP}. The challenges associated with the LMMSE channel estimation algorithm are as follows:

$\bullet$ Insufficient orthogonal pilots ($0<\tau\ll M$): Due to the limited data in the observed signal $\mathbf{y}$, the empirical information contained in the second term of \eqref{eq8} is also highly constrained. The correction capability of this term for $\overline{\mathbf{h}}$ diminishes with a decreasing number of pilot signals. Consequently, the channel estimation performance deteriorates compared to scenarios with sufficient pilot.

$\bullet$ Inadequate SNR: In this case, the second term in \eqref{eq8} approaches zero, thereby yielding $\widehat{\mathbf{h}}_{\text{LMMSE}}\approx\overline{\mathbf{h}}$. Evidently, as the SNR decreases, the channel estimation error stabilizes at a constant value.

$\bullet$ Sophisticated groundtruth channel statistical distribution: The LMMSE algorithm utilizes only the first-order (mean) and second-order (covariance) statistical properties of the channel $\mathbf{h}$. When $P_h\left(\mathbf{h}\right)$ follows Gaussian distribution, the LMMSE algorithm can achieve an optimal solution. However, for more sophisticated groundtruth channel statistical distribution, the channel estimation results are generally suboptimal under the criterion of minimizing the MSE.

$\bullet$ No pilot signal ($\tau=0$): In this case, $\mathbf{X}=\mathbf{0}$. For equation \eqref{eq8}, this leads to $\widehat{\mathbf{h}}_{\text{LMMSE}}=\overline{\mathbf{h}}$. Therefore, the error of channel estimation depends on the covariance matrix $\mathbf{C}_{\mathbf{h}}$.

\subsubsection{MMSE Channel Estimation}
The fundamental principle of the MMSE channel estimation algorithm is to minimize the MSE between the estimated channel $\widehat{\mathbf{h}}_{\rm MMSE}$ and the groundtruth channel $\mathbf{h}$. The channel estimation result can be expressed as \cite{SSP}
\begin{equation}
\begin{split}
\widehat{\mathbf{h}}_{\rm MMSE} &= \mathbb{E}\left\{\mathbf{h}|\mathbf{y}\right\}=\int \mathbf{h}P_{h|y}\left(\mathbf{h}|\mathbf{y}\right)\,d\mathbf{h}\\
&=\frac{\int \mathbf{h}P_{y|h}\left(\mathbf{y}|\mathbf{h}\right)P_h\left(\mathbf{h}\right)\,d\mathbf{h}}{P_y\left(\mathbf{y}\right)}, \label{eq9}
\end{split}
\end{equation}
where $P_{h|y}\left(\mathbf{h}|\mathbf{y}\right)$, $P_h\left(\mathbf{h}\right)$ and $P_y\left(\mathbf{y}\right)$ are the posterior distribution of the channel $\mathbf{h}$, the PDF of channel $\mathbf{h}$, and the PDF of $\mathbf{y}$, respectively. 

Specifically, when $P_h\left(\mathbf{h}\right)$ follows Gaussian distribution, $\widehat{\mathbf{h}}_{\rm MMSE}=\widehat{\mathbf{h}}_{\rm LMMSE}$, with the specific expression given in equation \eqref{eq8}. And the challenges faced by the MMSE channel estimation algorithm are as follows:

$\bullet$ Insufficient orthogonal pilots ($0<\tau\ll M$): Due to the extremely limited information in the observation signal $\mathbf{y}$, the posterior distribution $P_{h|y}\left(\mathbf{h}|\mathbf{y}\right)$ provides less effective information. Compared to scenarios with sufficient pilot, the performance of the MMSE algorithm deteriorates.

$\bullet$ Inadequate SNR: In this case, $\mathbf{y}\approx\mathbf{z}$, implying $P_y\left(\mathbf{y}\right)\approx P_{z}\left(\mathbf{z}\right)=P_{y|h}\left(\mathbf{y}|\mathbf{h}\right)$. Consequently, according to equation \eqref{eq9}, $\widehat{\mathbf{h}}_{\text{MMSE}}\approx\mathbb{E}\left\{\mathbf{h}\right\}=\overline{\mathbf{h}}$. Evidently, as the SNR decreases, the channel estimation error degenrates to the mean.

$\bullet$ Sophisticated groundtruth channel statistical distribution: As analyzed above, when $P_h\left(\mathbf{h}\right)$ follows simple distributions such as Gaussian distribution, closed-form expressions for channel estimation can be directly obtained. However, for sophisticated groundtruth channel statistical distribution, the MMSE algorithm faces two main challenges: First, it is difficult to precisely obtain $P_{h|y}\left(\mathbf{h}|\mathbf{y}\right)$ (or $P_h\left(\mathbf{h}\right)$, $P_y\left(\mathbf{y}\right)$ and $P_{y|h}\left(\mathbf{y}|\mathbf{h}\right)$). Second, even if these statistical distributions are obtained, the integration operation required in \eqref{eq9} is difficult to perform. 

$\bullet$ No pilot signal ($\tau=0$): In this case, $P_{y|h}\left(\mathbf{y}|\mathbf{h}\right)=P_y\left(\mathbf{y}\right)=P_{z}\left(\mathbf{z}\right)$. According to equation \eqref{eq9}, $\widehat{\mathbf{h}}_{\text{MMSE}}=\mathbb{E}\left\{\mathbf{h}\right\}=\overline{\mathbf{h}}$. Again, the generation quality is poor if the groundtruth channel has a large covariance. 

\subsubsection{MAP Channel Estimation}
The MAP estimator estimates the channel by selecting the parameter values that maximize the posterior probability $P_{h|y}\left(\mathbf{h}|\mathbf{y}\right)$. Employing Bayes rule, this translates the estimation endeavor into an optimization problem of the following form
\begin{equation}
\begin{split}
\widehat{\mathbf{h}}_{\text{MAP}}&=\underset{\mathbf{h}}{\mathop{\text{arg}\max }}\,P_{h|y}\left(\mathbf{h}|\mathbf{y}\right)=\underset{\mathbf{h}}{\mathop{\text{arg}\max }}\,\frac{P_{y|h}\left(\mathbf{y}|\mathbf{h}\right)P_h\left(\mathbf{h}\right)}{P_y\left(\mathbf{y}\right)}\\
&=\underset{\mathbf{h}}{\mathop{\text{arg}\min }}\,-\text{log}\{P_{y|h}\left(\mathbf{y}|\mathbf{h}\right)\}-\text{log}\{P_h\left(\mathbf{h}\right)\}\\
&=\underset{\mathbf{h}}{\mathop{\text{arg}\min }}\,\frac{1}{2\sigma^{2}}\left|\left|\sqrt{\rho\xi}\mathbf{Xh}-\mathbf{y}\right|\right|^{2}-\text{log}\{P_h\left(\mathbf{h}\right)\}.\label{eq10}
\end{split}
\end{equation}
The challenges of the MAP channel estimation algorithm are as follows:

$\bullet$ Insufficient orthogonal pilots ($0<\tau\ll M$): According to equation \eqref{eq10}, the performance of the MAP channel estimation algorithm is jointly determined by the pilot observation term $\frac{1}{2\sigma^{2}}\left|\left|\sqrt{\rho\xi}\mathbf{Xh}-\mathbf{y}\right|\right|^{2}$ and the channel statistical distribution term $\text{log}\left\{P_h\left(\mathbf{h}\right)\right\}$. With fewer pilot signals, the observation term provides less useful information, thereby weakening its ability to correct the channel prior term. Consequently, the performance degrades compared to scenarios with sufficient pilot resources.

$\bullet$ Inadequate SNR: In this case, the first term in equation \eqref{eq10} approaches zero, thereby yielding $\widehat{\mathbf{h}}_{\text{MAP}}\approx\underset{\mathbf{h}}{\mathop{\text{arg}\max }}\,P_h\left(\mathbf{h}\right)$. Evidently, as the SNR decreases, the channel estimation degenrates to the channel that gives the highest prior probability.

$\bullet$ Sophisticated groundtruth channel statistical distribution: The primary challenges of the MAP algorithm include the acquisition of accurate $P_h\left(\mathbf{h}\right)$ (or $P_{h|y}\left(\mathbf{h}|\mathbf{y}\right)$) and the efficient solution of equation \eqref{eq10}.

$\bullet$ No pilot signal ($\tau=0$): According to equation \eqref{eq10}, $\widehat{\mathbf{h}}_{\text{MAP}}=\underset{\mathbf{h}}{\mathop{\text{arg}\max }}\,P_h\left(\mathbf{h}\right)$. However, high-quality channel estimation is only achievable when $P_h\left(\mathbf{h}\right)$ has little uncertainty, say in the extreme case where $P_h\left(\mathbf{h}\right)$ is a Dirac delta function.

Based on the above analysis, the LMMSE, MMSE, and MAP algorithms, which incorporate the statistical properties of the channel, are anticipated to outperform the LS and ML algorithms in scenarios with limited orthogonal pilot signals, low SNR, or no pilot signals. Compared to the LMMSE algorithm, the MMSE and MAP algorithms use complete channel prior distribution information, thus they are expected to achieve better channel estimation performance in any complex channel distribution scenario. Furthermore, compared to the MMSE algorithm, the MAP algorithm leverages the application of Bayes rule to simplify its computational process. It only requires the acquisition of the channel PDF $P_h\left(\mathbf{h}\right)$ and subsequently employs optimization algorithms to solve the optimization problem \eqref{eq10}. Unlike MMSE channel estimation, it does not require obtaining the complete channel posterior distribution and perform sophisticated integration. Additionally, the MMSE channel estimation algorithm is known to suffer from the ``regression to the mean'' issue, in which case the inferred average solution may lack important details or even exceed the expected solution space \cite{InverseProblem}, while the MAP algorithm does not have this problem.

\section{Proposed CKM-Based Channel Estimation}
Based on the above analysis, this paper focuses on the MAP channel estimation method, emphasizing the acquisition of complex channel priors $P_h\left(\mathbf{h}\right)$ and the efficient handling of the optimization problem \eqref{eq9}. Besides, it is useful to have an additional regularization parameter for the MAP method to control the relative effect of the prior information on the channel estimation. Therefore, the channel estimation problem studied in this paper can be formulated as follows
\begin{equation}
\widehat{\mathbf{h}}_{\text{rMAP}}=\underset{\mathbf{h}}{\mathop{\text{arg}\min }}\,\frac{1}{2\sigma^{2}}\left|\left|\sqrt{\rho\xi}\mathbf{Xh}-\mathbf{y}\right|\right|^{2}-\beta \text{log}\{P_h\left(\mathbf{h}\right)\},\label{eq11}
\end{equation}
where $\beta>0$ is the regularization parameter. Specifically, when $P_h\left(\mathbf{h}\right)$ is a simple distribution such as Gaussian, uniform, or impulse, or a complex distribution that is differentiable, problem \eqref{eq11} can be directly solved.

$\bullet$ When $P_h\left(\mathbf{h}\right)$ is Gaussian distribution, the channel estimation is given by
\begin{equation}
\begin{split}
\widehat{\mathbf{h}}_{\text{rMAP,G}}=&\overline{\mathbf{h}}+\beta\sqrt{\rho\xi}\mathbf{C}_{\mathbf{h}}\mathbf{X}^H\left(\rho\xi\mathbf{X}\mathbf{C}_{\mathbf{h}}\mathbf{X}^H+\sigma^2\mathbf{I}_{\tau}\right)^{-1}\cdot\\
&\left(\mathbf{y}-\sqrt{\rho\xi}\mathbf{X}\overline{\mathbf{h}}\right).\label{eq12}
\end{split}
\end{equation}

$\bullet$ When $P_h\left(\mathbf{h}\right)$ is uniform distribution, it is constant for all $\mathbf{h}$, rendering the second term in \eqref{eq11} ineffective for channel estimation. Thus, $\widehat{\mathbf{h}}_{\text{rMAP,U}}=\widehat{\mathbf{h}}_{\text{LS}}$.

$\bullet$ When $P_h\left(\mathbf{h}\right)$ is a Dirac delta function centered at $\mathbf{h}_{D}$, the channel estimate is simply $\widehat{\mathbf{h}}_{\text{rMAP,D}}=\mathbf{h}_{D}$.

$\bullet$ When $P_h\left(\mathbf{h}\right)$ follows a distribution that is explicitly expressed and differentiable, problem \eqref{eq11} can be solved via classical gradient descent methods. For example, the iterative procedure of the SD method \cite{Convexoptimization} is
\begin{equation}
\widehat{\mathbf{h}}_{i+1}=\widehat{\mathbf{h}}_{i}-\delta_i\nabla_{\mathbf{h}}\left(\frac{\left|\left|\sqrt{\rho\xi}\mathbf{Xh}-\mathbf{y}\right|\right|^{2}}{2\sigma^{2}}-\beta \text{log}\{P_h\left(\mathbf{h}\right)\} \right)\bigg|_{\mathbf{h}=\widehat{\mathbf{h}}_i},\label{eq13}
\end{equation}
where the subscript $i$ is the iteration index, and $\delta_i$ is the step size in the $i$-th iteration.

However, when $P_h\left(\mathbf{h}\right)$ is a complex distribution that is non-differentiable, the optimization problem \eqref{eq11} becomes intractable for direct solution. In such cases, the PnP \mbox{technique \cite{PnP}} is employed to simplify the optimization problem \eqref{eq11}. Subsequently, the CKM technique \cite{CKMTutorial, CKMMagazine} is utilized to efficiently process the channel prior distribution term, thereby achieving low-complexity and high-precision channel estimation. 

\subsection{PnP for Problem Simplification}
For the optimization problem \eqref{eq11} that incorporates a complex and non-differentiable $P_h\left(\mathbf{h}\right)$, the core processing procedure of the PnP technique consists of two steps. Firstly, by employing classical variable decomposition algorithms such as the alternating direction method of multipliers (ADMM) \cite{ADMM} and the half-quadratic splitting \mbox{(HQS) \cite{HQS}}, the observation term (first term) and regularization term (second term) in equation \eqref{eq11} are decoupled, thereby transforming equation \eqref{eq11} into two subproblems. Subsequently, the two subproblems are optimized independently and alternated iteratively until convergence, resulting in the estimated channel.

For the first step, the channel vector $\mathbf{h}$ is decomposed into two new vectors $\mathbf{h}$ and $\mathbf{v}$. Consequently, the optimization problem in \eqref{eq11} can be reformulated as follows
\begin{equation}
\begin{split}
\widehat{\mathbf{h}}_{\text{rMAP-PnP}}=\underset{\mathbf{h}}{\mathop{\text{arg}\min }}\,\frac{1}{2\sigma^{2}}&\left|\left|\sqrt{\rho\xi}\mathbf{Xh}-\mathbf{y}\right|\right|^{2}-\beta \text{log}\left\{P_v\left(\mathbf{v}\right)\right\}\\
&\text{ s.t. } \mathbf{h}=\mathbf{v}. \label{eq14}
\end{split}
\end{equation}
The augmented Lagrangian for \eqref{eq14} is given by
\begin{equation}
L=\frac{1}{2\sigma^{2}}\left|\left|\sqrt{\rho\xi}\mathbf{Xh}-\mathbf{y}\right|\right|^{2}-\beta \text{log}\left\{P_v\left(\mathbf{v}\right)\right\}+\frac{\mu }{2}\left|\left|\mathbf{h}-\mathbf{v}\right|\right|^{2}, \label{eq15}
\end{equation}
where $\mu >0$ is the penalty parameter. The minimum of the optimization problem \eqref{eq15} corresponds to the saddle point of $L$ \cite{PnPConvergence}, which can be obtained by iteratively performing the following two steps until convergence.
\begin{equation}
\widehat{\mathbf{h}}_{i+1}\leftarrow  \underset{\mathbf{h}}{\mathop{\text{arg}\min }}\,\frac{1}{2\sigma^{2}}\left|\left|\sqrt{\rho\xi}\mathbf{Xh}-\mathbf{y}\right|\right|^{2}+\frac{\mu_{i}}{2}\left|\left|\mathbf{h}-\widehat{\mathbf{v}}_{i}\right|\right|^{2}, \label{eq16}
\end{equation}
\begin{equation}
\widehat{\mathbf{v}}_{i+1}\leftarrow \underset{\mathbf{v}}{\mathop{\text{arg}\min }}\, \frac{\mu_{i}}{2}\left|\left|\widehat{\mathbf{h}}_{i+1}-\mathbf{v}\right|\right|^{2}-\beta \text{log}\left\{P_v\left(\mathbf{v}\right)\right\}, \label{eq17}
\end{equation}
where the subscript $i$ represents the number of iterations. The minimization problem in formula \eqref{eq16} is a typical quadratic programming problem concerning the variable $\mathbf{h}$. Notably, since both $\sigma^{2}$, $\mu_{i}$, $\rho$ and $\xi$ are positive real values, the Hessian matrix associated with this problem is positive definite. Consequently, by computing the partial derivatives with respect to $\mathbf{h}$ and subsequently determining the extremum, the optimal value for problem \eqref{eq16} is
\begin{equation}
\widehat{\mathbf{h}}_{i+1} = \left(\rho\xi\mathbf{X}^{H} \mathbf{X}+\mu_{i}\sigma^{2} \mathbf{I}_M\right)^{-1}\left(\sqrt{\rho\xi}\mathbf{X}^{H} \mathbf{y}+\mu_{i} \sigma^{2}\widehat{\mathbf{v}}_{i}\right). \label{eq18}
\end{equation}

On the other hand, the complex channel distribution $P_h\left(\mathbf{h}\right)$ investigated in this paper is typically non-differentiable and often lacks a closed-form expression. Consequently, the sub-optimization problem \eqref{eq17} cannot be directly solved using the classical SD algorithm and instead necessitates leveraging the learning capability of CKM for handling the complex $P_h\left(\mathbf{h}\right)$.

\subsection{CKM: Environment-Aware Channel Prior}
CKM, proposed in \cite{CKMTutorial} and \cite{CKMMagazine}, can efficiently provide channel statistical prior, and serves as a mapping function $\mathcal{M}$ from the location $\mathbf{q}$ to the channel PDF $P_{h|q}\left(\mathbf{h}|\mathbf{q}\right)$, i.e.,
\begin{equation}
\mathcal{M}: \mathbf{q}\rightarrow P_{h|q}\left(\mathbf{h}|\mathbf{q}\right).\label{eq19}
\end{equation}

\begin{figure}[htbp]
    \centering
    \subfigure[$P_h\left(\mathbf{h}\right)$ for entire region.]{
        \includegraphics[width=0.42\textwidth]{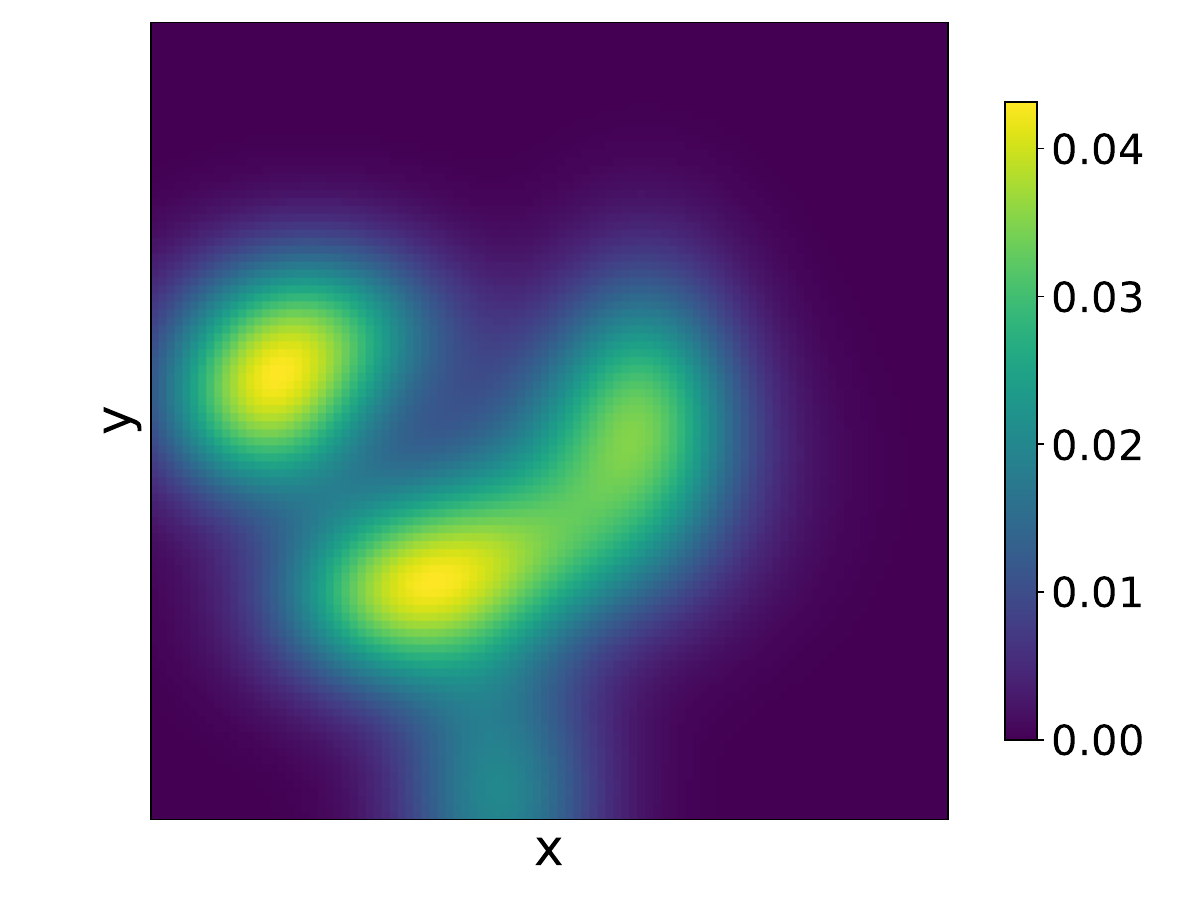}
        \label{figPDFEntireRegion}
    }
    \subfigure[$P_{h_p}\left(\mathbf{h}\right)$ for grid $p$.]{
        \includegraphics[width=0.42\textwidth]{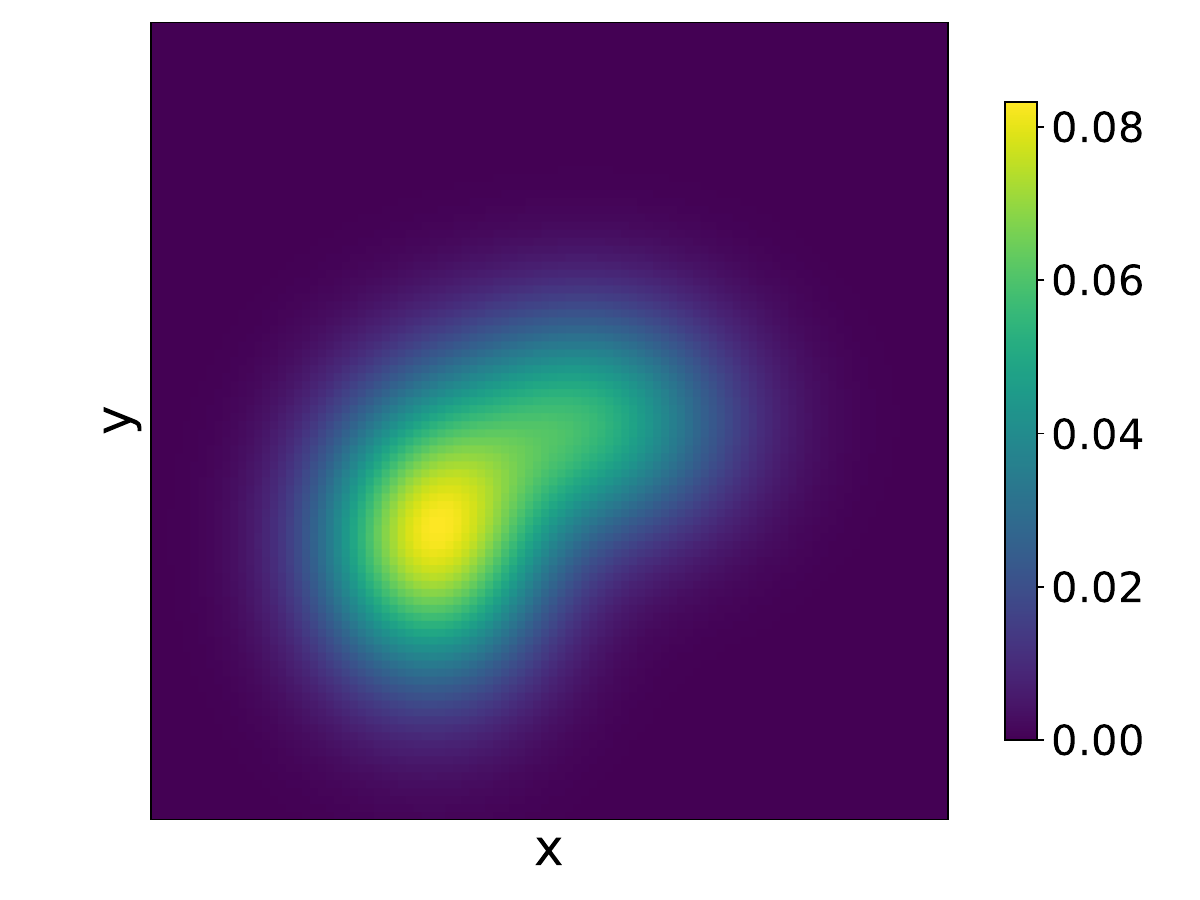}
        \label{figPDFSubRegion}
    }
    \caption{Schematic of channel distribution characteristics.}
    \label{figPDFRegion}
\end{figure}

CKM deviates from conventional methods of acquiring channel prior knowledge by explicitly emphasizing the strong coupling between wireless channel characteristics and environmental features at specific location $\mathbf{q}$. This approach effectively reduces the uncertainty associated with $P_h\left(\mathbf{h}\right)$ and has potential for efficiently addressing the optimization problem in \eqref{eq17}. For example, in the wireless communication environment, the channel prior distribution across the entire target area is typically coarse, leading to high uncertainty in the channel distribution, as illustrated in Fig. \ref{figPDFEntireRegion}. In contrast, by discretizing the entire area into uniform grids of size $d$, the channel distribution within a random grid $p$ can be refined, significantly reducing the uncertainty of the channel distribution. The distribution characteristics of $P_{h_p}\left(\mathbf{h}\right)$ in this grid are shown in Fig. \ref{figPDFSubRegion}. This improvement arises because each grid encompasses fewer environment features, allowing for more precise capture of their impact on the wireless channel, ultimately yielding a more accurate channel prior distribution. Therefore, to obtain more precise channel statistical prior information, it is necessary to discretize the entire target area.

In complex environments, closed-form expressions for channel statistical distributions are generally unavailable, and differentiation is impractical. Consequently, based on the aforementioned analysis, it is impossible to directly solve the problem \eqref{eq17} using various classical gradient descent algorithms. Fortunately, it is noted that the partial derivative of $\text{log}\left\{P_h\left(\mathbf{h}\right)\right\}$ with respect to $\mathbf{h}$ is known as the score \mbox{function \cite{ScoreFunction}}. According to Tweedie's formula \cite{Tweedie}, it has the following theorem with the MMSE estimator:

\begin{theorem}
Consider an observation $\widetilde{\mathbf{h}}$, which is generated from the clean channel $\mathbf{h}\sim P_h\left(\mathbf{h}\right)$ by additive Gaussian noise, i.e., $\widetilde{\mathbf{h}}=\mathbf{h}+\widetilde{\sigma}\boldsymbol{\epsilon}$, where $\boldsymbol{\epsilon}$ follows a standard Gaussian distribution with mean zero and covariance matrix being the identity matrix. Define the marginal distribution $P_{\tilde{h}}\left(\widetilde{\mathbf{h}}\right)=\int P_{\tilde{h}|h}\left(\widetilde{\mathbf{h}}|\mathbf{h}\right)P_h\left(\mathbf{h}\right)\, d\mathbf{h}$. Then, its score function $\nabla_{\widetilde{\mathbf{h}}}\left(\text{log}\left\{P_{\tilde{h}}\left(\widetilde{\mathbf{h}}\right)\right\} \right)$ and the MMSE denoiser $\mathbb{E}\left\{\mathbf{h}|\widetilde{\mathbf{h}}\right\}$ satisfy the following relationship
\begin{equation}
\nabla_{\widetilde{\mathbf{h}}}\left(\text{log}\left\{P_{\tilde{h}}\left(\widetilde{\mathbf{h}}\right)\right\} \right)=\frac{1}{\widetilde{\sigma}^2}\left(\mathbb{E}\left\{\mathbf{h}|\widetilde{\mathbf{h}}\right\}-\widetilde{\mathbf{h}}\right).\label{eq20}
\end{equation}
\label{Theorem1}
\end{theorem}

It is worth noting that Theorem \ref{Theorem1} imposes no restrictions on the distribution of the groundtruth channel $\mathbf{h}$, thus rendering it applicable to the problem investigated in this paper. However, in practical applications, the MMSE estimator $\mathbb{E}\left\{\mathbf{h}|\widetilde{\mathbf{h}}\right\}$ is often difficult to obtain directly. An effective solution is to leverage the advanced data-processing capabilities of AI to train a denoiser $D\left(\widetilde{\mathbf{h}},\widetilde{\sigma}\right)$ that approximates $\mathbb{E}\left\{\mathbf{h}|\widetilde{\mathbf{h}}\right\}$ \cite{DenoiserMMSE}. 

This paper proposes a novel CKM, namely CSFM, to provide high-quality channel score functions for each location within the target area, thereby effectively addressing the problem described in \eqref{eq17}. The CSFM primarily stores the architecture and parameters of the channel denoiser neural network, which directly influences the accuracy of the score function. Therefore, constructing a CSFM for the entire area using massive historical channel data and AI techniques may face significant challenges due to the large scale of the area. Specifically, the complex and diverse physical environmental features within the area significantly impact the channel distribution, potentially requiring deeper network architectures and more historical channel data for effective learning. This poses substantial demands on computational and temporal resources. To address these challenges, this paper discretizes the target area using a uniform grid of size $d$, and trains a channel denoiser based on the historical channel data within each grid. For each grid after discretization, the CSFM stores the corresponding denoiser neural network architecture and parameters. Moreover, to maximize the provision of channel prior information within limited storage space, the CSFM selectively retains sparse channel data at discrete locations from the massive historical channel data.

Based on the above analysis, the channel denoiser provided by the CSFM can approximate the channel score function. Consequently, the optimization problem \eqref{eq17} can be solved using the SD method, with the expression given as follows:
\begin{equation}
\widehat{\mathbf{v}}_{i+1}=\widehat{\mathbf{v}}_{i}-\delta_i\nabla_{\mathbf{v}}\bigg(\frac{\mu_{i}}{2}\left|\left|\widehat{\mathbf{h}}_{i+1}-\mathbf{v}\right|\right|^{2}-\beta \text{log}\{P_v\left(\mathbf{v}\right)\}\bigg)\bigg|_{\mathbf{v}=\widehat{\mathbf{v}}_{i}}, \label{eq21}
\end{equation}
where $\delta_i$ is the step size at the $i$-th iteration. Specifically,
\begin{equation}
\nabla_{\mathbf{v}}\left(\frac{\mu_{i}}{2}\left|\left|\widehat{\mathbf{h}}_{i+1}-\mathbf{v}\right|\right|^{2}\right)\bigg|_{\mathbf{v}=\widehat{\mathbf{v}}_{i}}=\mu_{i}\left(\widehat{\mathbf{v}}_{i}-\widehat{\mathbf{h}}_{i+1}\right).\label{eq22}
\end{equation}
Besides, according to the Theorem \ref{Theorem1}, the following equation can be obtained
\begin{equation}
\nabla_{\mathbf{v}}\left(-\beta \text{log}\{P_v\left(\mathbf{v}\right)\} \right)\big|_{\mathbf{v}=\widehat{\mathbf{v}}_{i}}=\frac{\beta}{\widetilde{\sigma}_{i}^{2}}\left(\widehat{\mathbf{v}}_{i}-D\left(\widehat{\mathbf{v}}_{i},\widetilde{\sigma}_{i}\right)\right), \label{eq23}
\end{equation}
where $D\left(\cdot,\cdot\right)$ is the channel denoiser, whose inputs are the noisy channel $\widehat{\mathbf{v}}_{i}$ and the variance of the noise to be removed, i.e., $\widetilde{\sigma}_{i}^{2}$. Therefore, based on equations \eqref{eq22} and \eqref{eq23}, the solution for \eqref{eq21} can be expressed as follows
\begin{equation}
\widehat{\mathbf{v}}_{i+1}=\widehat{\mathbf{v}}_{i}-\delta_i\left(\mu_{i}\left(\widehat{\mathbf{v}}_{i}-\widehat{\mathbf{h}}_{i+1}\right)+\frac{\beta}{\widetilde{\sigma}_{i}^{2}}\left(\widehat{\mathbf{v}}_{i}-D\left(\widehat{\mathbf{v}}_{i},\widetilde{\sigma}_{i}\right)\right)\right). \label{eq24}
\end{equation}

To simplify the parameter selection in equation \eqref{eq24}, according to formula \eqref{eq19}, Venkatakrishnan \textit{et al.} \cite{PnP, DDUNet} proposed that $\widetilde{\sigma}_{i}$ can be directly denoted as $\sqrt{\beta / \mu_{i}}$. On the other hand, due to the unknown partial derivative of $D\left(\widehat{\mathbf{v}}_{i},\widetilde{\sigma}_{i}\right)$ with respect to $\widehat{\mathbf{v}}_{i}$, the direct computation of the optimal step size $\delta_i$ is not feasible. Under the condition of $\widetilde{\sigma}_{i}=\sqrt{\beta / \mu_{i}}$, it is possible to set $\delta_i=\frac{\alpha}{\mu_{i}}$. This simplification effectively transforms the selection of multiple parameters in equation \eqref{eq24} into the determination of a single parameter $\alpha$.

\subsection{CSFM-PnP Based Channel Estimation}
The proposed CSFM-PnP based environment-aware channel estimation algorithm, leveraging SCFM and PnP techniques, can effectively address the optimization problem \eqref{eq11} to achieve low-overhead, high-precision channel estimation in complex wireless environments. This paper focuses on channel estimation within an arbitrary grid $p$, where the estimation process at location $\mathbf{q}$ within grid $p$ involves rewriting the variables in equations \eqref{eq18} and \eqref{eq24} $\widehat{\mathbf{h}}_i\left(\mathbf{q}\right)$, $\widehat{\mathbf{v}}_i\left(\mathbf{q}\right)$, $\xi\left(\mathbf{q}\right)$, $\mu_{i}\left(\mathbf{q}\right)$, $\sigma^{2}\left(\mathbf{q}\right)$, $\delta_i\left(\mathbf{q}\right)$, and $D_p\left(\cdot,\cdot\right)$. Specifically, the $D_p\left(\cdot,\cdot\right)$ is the channel denoiser trained by the historical channel data in grid $p$. Besides, based on the above analysis, by iteratively solving equation \eqref{eq18} and \eqref{eq24} until convergence, the final channel estimation result can be obtained. However, during this process, two new issues arise. The first issue pertains to ensuring the convergence of the iterative process, while the second issue involves the selection of appropriate initial values to enhance the accuracy of channel estimation.

The first issue can be effectively addressed by progressively increasing the penalty coefficient $\mu_i\left(\mathbf{q}\right)$ in equation \eqref{eq18} and \eqref{eq19} throughout the iterative process \cite{PnPConvergence}. One of the classic and straightforward methods is
\begin{equation}
\mu_{i+1}\left(\mathbf{q}\right)=\gamma\cdot\mu_{i}\left(\mathbf{q}\right), \label{eq25}
\end{equation}
where $\gamma> 1$ is a constant. At the same time, note that $\widetilde{\sigma}_{i}\left(\mathbf{q}\right)=\sqrt{\beta / \mu_{i}\left(\mathbf{q}\right)}$, which implies that for a given regularization coefficient $\beta$, $\widetilde{\sigma}_{i}\left(\mathbf{q}\right)$ will continuously decrease with the increase in the number of iterations. This aligns with the concept of gradual denoising, where initially, a larger noise magnitude is input into the denoiser for a coarser denoising, and as the iterative process progresses, increasingly refined denoising is performed.

The second issue is the selection of initial values for $\widehat{\mathbf{h}}_{0}\left(\mathbf{q}\right)$, $\widehat{\mathbf{v}}_{0}\left(\mathbf{q}\right)$, $\mu_{0}\left(\mathbf{q}\right)$, $\beta$, $\gamma$ and $\alpha$. These values can be classified into two categories, each with a corresponding initial value selection scheme. $\widehat{\mathbf{h}}_{0}\left(\mathbf{q}\right)$ and $\widehat{\mathbf{v}}_{0}\left(\mathbf{q}\right)$ pertain to the issue of selecting initial values related to channel parameters. Conversely, $\mu_{0}\left(\mathbf{q}\right)$, $\beta$, $\gamma$ and $\alpha$, concern the parameter selection for the PnP algorithm. 

For the challenging scenarios considered in this paper, the traditional methods for selecting initial values $\widehat{\mathbf{h}}_{0}\left(\mathbf{q}\right)$ and $\widehat{\mathbf{v}}_{0}\left(\mathbf{q}\right)$ are often insufficient, as both the direct utilization of the observation signal $\mathbf{y}$ and the LS channel estimate will be unsuitable. To address this issue, the CSFM stores historical channel information at discrete points within the target area, providing environment-aware initial estimates to enhance the iterative process. By leveraging the distances between the location $\mathbf{q}$ of the UE and the discrete sampling points in the CSFM, an efficient acquisition of $\widehat{\mathbf{h}}_\text{CSFM-NN}\left(\mathbf{q}\right)$ can be achieved based on the nearest neighbor (NN) algorithm. Then, $\widehat{\mathbf{h}}_{0}\left(\mathbf{q}\right)$ and $\widehat{\mathbf{v}}_{0}\left(\mathbf{q}\right)$ can be set as $\widehat{\mathbf{h}}_\text{CSFM-NN}\left(\mathbf{q}\right)$.

The selection of $\mu_{0}\left(\mathbf{q}\right)$, $\beta$, $\gamma$ and $\alpha$ is based on coordinate descent. To reduce complexity,  $\gamma$ is initially fixed at $2$. The remaining parameters are optimized by fixing any two of the variables $\mu_{0}\left(\mathbf{q}\right)$, $\beta$ and $\alpha$ while optimizing the third. This iterative approach gradually improves performance towards an optimal state. Additionally, equation \eqref{eq20} guides the selection of $\mu_{0}\left(\mathbf{q}\right)$. To balance the influence of the observed signal $\mathbf{y}\left(\mathbf{q}\right)$ and the denoised channel $\mathbf{v}_{i}\left(\mathbf{q}\right)$ on the update of $\mathbf{h}_{i+1}\left(\mathbf{q}\right)$, $\mu_{i}\left(\mathbf{q}\right) \sigma^{2}\left(\mathbf{q}\right)$ is set to match the magnitude of the non-zero diagonal elements of $\rho\xi\left(\mathbf{q}\right)\mathbf{X}^{H}\mathbf{X}$. Thus, $\mu_{0}\left(\mathbf{q}\right)$ is set as $\frac{\alpha^{\prime}\rho\xi\left(\mathbf{q}\right)}{\sigma^{2}\left(\mathbf{q}\right)}$, simplifying the search to values of $\alpha^{\prime}$ within a small range.

Informed by the above analysis, the environment-aware channel estimation based on CSFM-PnP can be organized into Algorithm \ref{alg1} as follows.
\begin{algorithm}[h]
\caption{CSFM-PnP-Based Environment-Aware Channel Estimation Algorithm.}
\label{alg1}
\begin{algorithmic}[1]
\Statex \textbf{Input:} CSFM, UE location $\mathbf{q}$, observed signal $\mathbf{y}\left(\mathbf{q}\right)$, noise variance $\sigma^{2}\left(\mathbf{q}\right)$, the normalized channel coefficient $\xi\left(\mathbf{q}\right)$, transmitted signal power $\rho$, pilot sequence $\mathbf{X}$, initial penalty coefficient $\mu_{0}\left(\mathbf{q}\right)$ (or $\alpha^{\prime}$), step size coefficient $\alpha$, regularization coefficient $\beta$, number of iterations $I$

\State \textbf{Initialization:} $\widehat{\mathbf{h}}_{0}\left(\mathbf{q}\right)=\widehat{\mathbf{v}}_{0}\left(\mathbf{q}\right)=\widehat{\mathbf{h}}_\text{CSFM-NN}\left(\mathbf{q}\right)$
\For{$i=0$ to $I-1$}
	\State Update $\widehat{\mathbf{h}}_{i+1}\left(\mathbf{q}\right)$ using equation \eqref{eq18}
	\State$\widetilde{\sigma}_{i}^{2}\left(\mathbf{q}\right)=\frac{\beta}{\mu_{i}\left(\mathbf{q}\right)}$
	\State$\delta_{i}\left(\mathbf{q}\right)=\frac{\alpha}{\mu_{i}\left(\mathbf{q}\right)}$
	\State Update $\widehat{\mathbf{v}}_{i+1}\left(\mathbf{q}\right)$ using equation \eqref{eq24}
	\State Update $\mu_{i+1}\left(\mathbf{q}\right)$ using equation \eqref{eq25}
\EndFor
\Statex \textbf{Output:} $\widehat{\mathbf{h}}_\text{CSFM-PnP}\left(\mathbf{q}\right)=\widehat{\mathbf{h}}_{I}\left(\mathbf{q}\right)$ 
\end{algorithmic}
\end{algorithm}

\section{Deep Denoising U-Net Based CSFM Construction}
This section focuses on the construction of the CSFM based on the deep denoising architecture, providing a detailed exposition of key aspects such as channel data preprocessing and network architecture design. 

\subsection{Construction of CSFM}
\begin{figure}[htbp]
\centering
\includegraphics[scale=0.37]{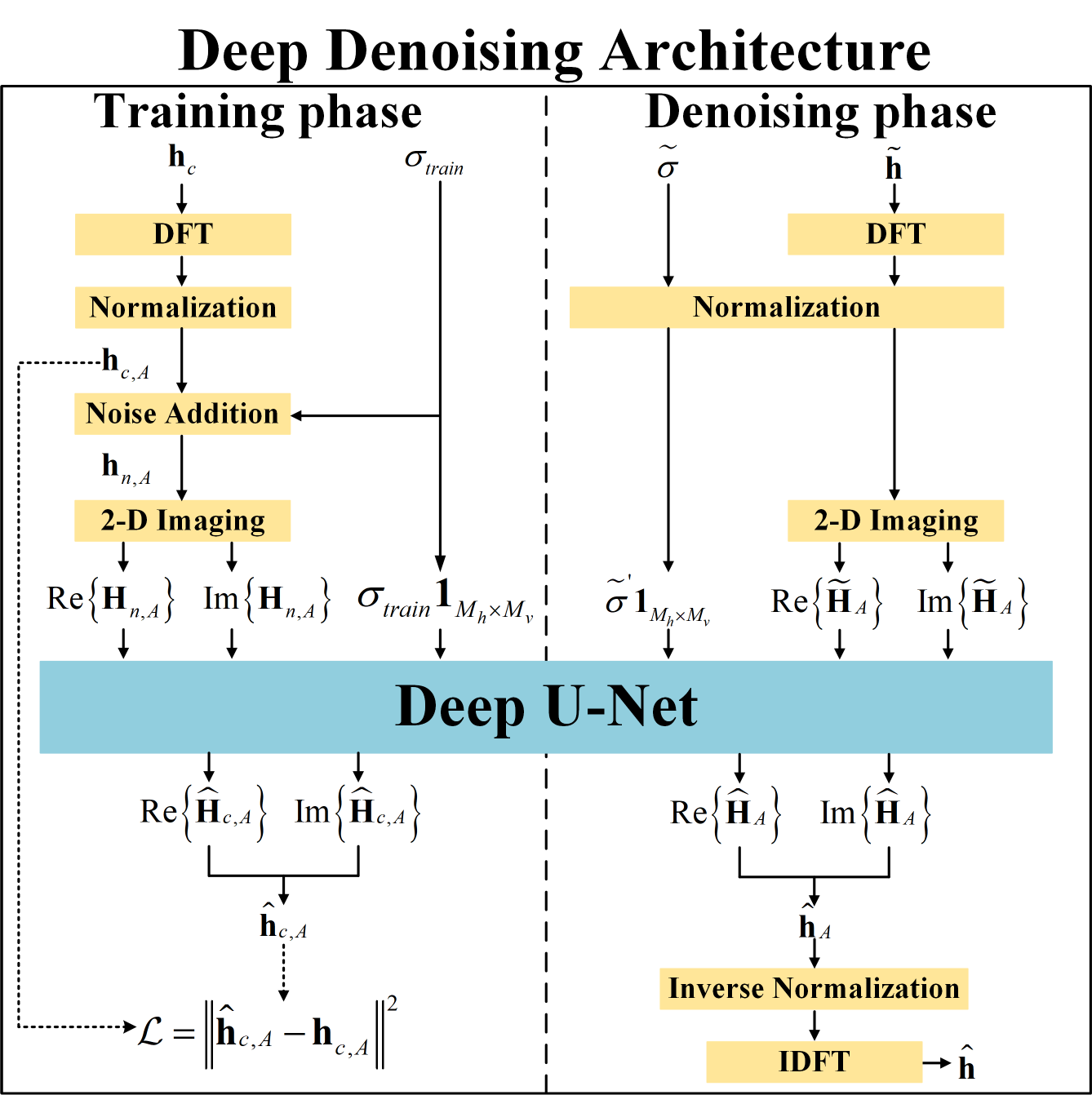}
\caption{Overall deep denoising architecture for denoiser training and channel denoising.}
\label{figDDUNet}
\end{figure}
The construction of channel denoisers in CSFM primarily draws inspiration from image denoising networks in the field of image processing. Fig. \ref{figDDUNet} provides a comprehensive overview of the training and application framework of the channel denoiser within the CSFM, with the core being a Deep U-Net, which will be elaborated in detail in subsequent subsections. The following are introductions to other modules in the framework:

\subsubsection{DFT/IDFT Module}
For the grid $p$ considered in this paper, the noise-free historical channel $\mathbf{h}_{c}\in \mathbb{C}^{M\times 1}$ is initially transformed into the angular domain by left-multiplying it with the discrete fourier transform (DFT) matrix \mbox{$\mathbf{F}\in\mathbb{C}^{M\times M}$ \cite{DLChannelEstimation}}. This transformation leverages the sparsity of the angular domain channel to significantly reduce the number of feature parameters that each channel needs to learn, thereby alleviating the complexity of the training process. Similarly, in the denoising phase, the final channel estimate result can be obtained by left-multiplying the denoised angular domain channel with the inverse discrete fourier transform (IDFT) matrix $\mathbf{F}^{H}\in\mathbb{C}^{M\times M}$.

\subsubsection{(Inverse) Normalization Module}
To mitigate the risk of gradient explosion or vanishing in deep neural networks due to the large disparities among the dimensions of the angular domain channel data, the angular domain channel needs to be normalized. This normalization adjusts both the real and imaginary parts to the $\left[0,1\right]$ range. The normalization process is as follows: firstly, determine the maximum and minimum values of the real and imaginary parts of each angular domain channel. Then, normalize each dimension of the angular domain channel individually, resulting in the normalized angular domain channel $\mathbf{h}_{c,A}\in\mathbb{C}^{M\times 1}$. Similarly, in the denoising phase, the normalized angular domain channel after denoising requires the execution of inverse normalization processing.

\subsubsection{Noise Addition Module}
Gaussian noise is added to the normalized angular domain channel $\mathbf{h}_{c,A}\in\mathbb{C}^{M\times 1}$ to obtain the noisy channel, and this process can be expressed as
\begin{equation}
\mathbf{h}_{n,A}=\mathbf{h}_{c,A}+\sigma_{train}\boldsymbol{\epsilon}_{train} ,\label{eq26}
\end{equation}
where $\sigma_{train}$ is the standard deviation of noise, whose value is randomly selected within the range of $\left[0.00001,0.01\right]$ during each noise addition process, and $\boldsymbol{\epsilon}_{train}\sim\mathcal{CN}\left(\mathbf{0},\mathbf{I}_{M}\right)$. This approach is employed to ensure that the trained network can adapt to denoising requirements across different noise intensities.

\subsubsection{2-D Imaging Module}
To emulate the two-dimensional image processing procedure, this paper adopts the following approach to process the channel vector. Initially, the channel vector $\mathbf{h}_{n,A}\in\mathbb{C}^{M\times 1}$ is transformed into a two-dimensional matrix $\mathbf{H}_{n,A}\in\mathbb{C}^{M_{h}\times M_{v}}$ with $M=M_h \times M_v$, where $M_h$ and $M_v$ represent the number of pixels in the horizontal and vertical directions of the two-dimensional image, respectively. Subsequently, the real part $\text{Re}\left\{\mathbf{H}_{n,A}\right\}$ and the imaginary part $\text{Im}\left\{\mathbf{H}_{n,A}\right\}$ of $\mathbf{H}_{n,A}$ are extracted and input into the Deep U-Net as two images with pixel values $M_{h}\times M_{v}$. Furthermore, to enable the trained denoiser to effectively remove noise with a specified standard deviation, an extra image with dimensions $M_{h}\times M_{v}$ is generated. All pixel values of this image are set to $\sigma_{train}$ as specified by equation \eqref{eq26}. This image is used as the third input to the Deep U-Net.

After the three 2-D images $\text{Re}\left\{\mathbf{H}_{n,A}\right\}$, $\text{Im}\left\{\mathbf{H}_{n,A}\right\}$, and $\sigma_{train}\mathbf{1}_{M_{h}\times M_{v}}$ are fed into the Deep U-Net, the denoised values of $\text{Re}\left\{\mathbf{H}_{n,A}\right\}$ and $\text{Im}\left\{\mathbf{H}_{n,A}\right\}$ are produced, denoted as $\text{Re}\left\{\widehat{\mathbf{H}}_{c,A}\right\}$ and $\text{Im}\left\{\widehat{\mathbf{H}}_{c,A}\right\}$, respectively. By using $\text{Re}\left\{\widehat{\mathbf{H}}_{c,A}\right\}$ as the real part and $\text{Im}\left\{\widehat{\mathbf{H}}_{c,A}\right\}$ as the imaginary part, the output $\widehat{\mathbf{H}}_{c,A}\in\mathbb{C}^{M_{h}\times M_{v}}$ is reconstructed. Subsequently, $\widehat{\mathbf{H}}_{c,A}$ is reshaped to yield the denoised and normalized data in the angular domain, denoted as $\widehat{\mathbf{h}}_{c,A}\in\mathbb{C}^{M\times 1}$. The loss function is then calculated using $\left|\left|\widehat{\mathbf{h}}_{c,A}-\mathbf{h}_{c,A}\right|\right|^2$ to adjust the network parameters of the model.
\subsection{Network Architecture of Deep U-Net}
\begin{figure}[htbp]
\centering
\includegraphics[scale=0.57]{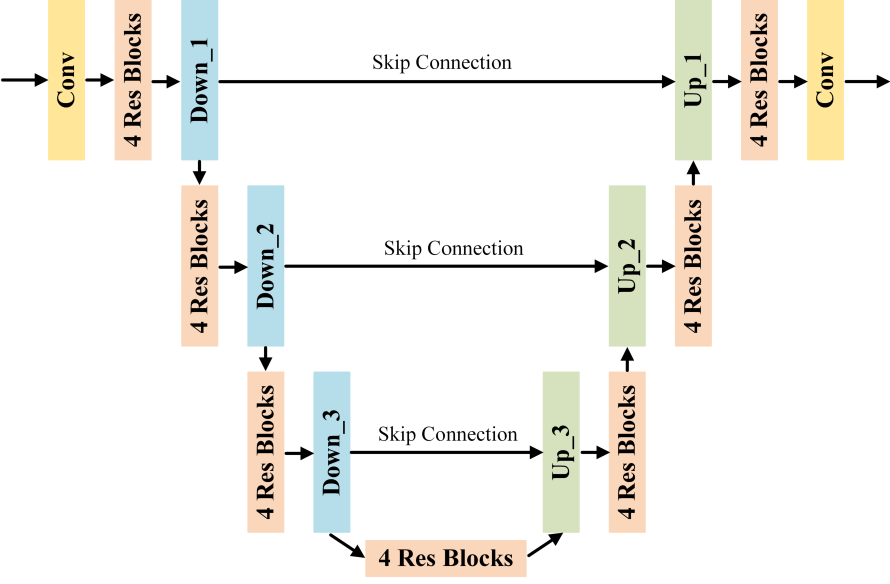}
\caption{The deep U-Net architecture integrates the U-Net and ResNet frameworks.}
\label{figUNet}
\end{figure}
Inspired by the image denoising research presented in \cite{DDUNet}, this paper employs a Deep U-Net-based denoising method to achieve efficient and high-quality noise removal for channel images $\text{Re}\left\{\mathbf{H}_{n,A}\right\}$ and $\text{Im}\left\{\mathbf{H}_{n,A}\right\}$. As illustrated in Fig. \ref{figUNet}, the Deep U-Net architecture comprises three downsampling and three upsampling operations, with adjacent downsampling or upsampling processes connected via a four-layer residual block. The Deep U-Net network integrates the exceptional performance of U-Net in the field of image-to-image \mbox{translation \cite{UNet}} and the unique advantages of ResNet in ensuring the stability and performance of deep network training \cite{ResNet}. Define the convolutional hyperparameters composed of kernel size, stride, and padding as (Kernel size, Stride, Padding). Specifically, each downsampling operation employs a strided convolution with the hyperparameters set to $\left(2,2,0\right)$. This operation effectively reduces both the height and width of each image by half. Similarly, in each upsampling operation, a transposed convolution with the convolutional hyperparameters specified as $\left(2,2,0\right)$ is applied, consequently leading to a twofold expansion in both the height and width dimensions of each image. Each downsampling process and its corresponding upsampling process are connected via an addition-based identity skip connection. Additionally, the convolutional hyperparameters used in both the ``Conv'' and the ``4 Res Blocks'' in Fig. \ref{figUNet} are set to $\left(3,1,1\right)$. This configuration ensures that the image pixel dimensions remain unchanged. And following the design outlined in \cite{DDUNet}, only each residual block contains one ReLU activation function.

\section{Simulation Results}
\subsection{Simulation Data Generation and Preprocessing}
In this section, the proposed CSFM-PnP-based environment-aware channel estimation algorithm is verified through ray-tracing simulation. The commercial ray-tracing software Remcom Wireless Insite\footnote{\url{https://www.remcom.com/wireless-insite-em-propagation-software}} is used to generate the groundtruth channel information. Then, the construction and application of CSFM are all based on this data. 
\begin{figure}[htbp]
\centering
\includegraphics[scale=0.45]{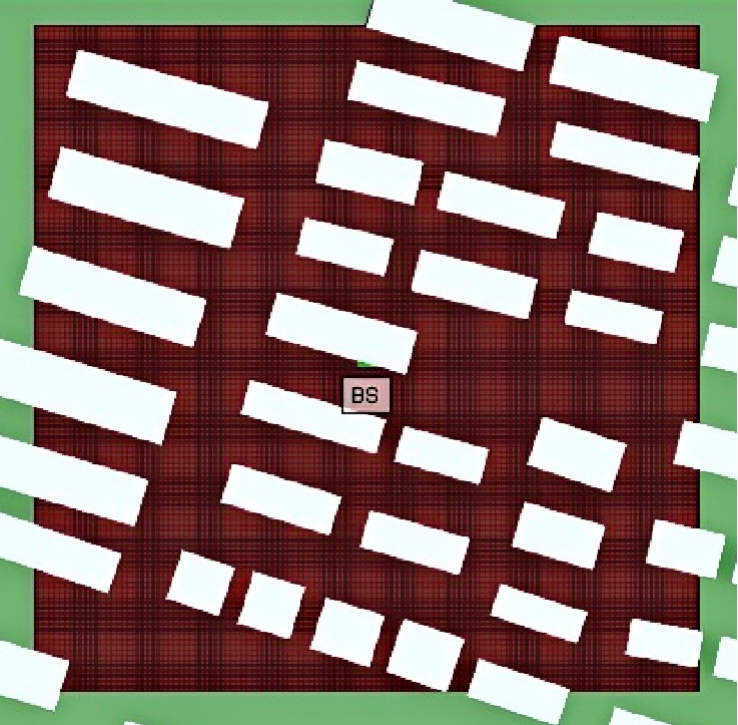}
\caption{The groundturth communication scenario in Wireless Insite.}
\label{figWIscenario}
\end{figure}

As illustrated in Fig. \ref{figWIscenario}, a $200\times 200$ square meters red area within a complex urban block is selected as the study scenario in this paper. The BS is positioned at the geometric center of this region. A total of $302959$ receiver points were uniformly distributed across the area to capture the channel characteristics at various locations, with a sampling interval of $\sqrt{2}/5$ meters. Furthermore, the BS was positioned at a height of $2$ meters above the ground. In this communication scenario, the BS employs a $16\times 16$ uniform planar array (UPA) to transmit signals at the $28$ GHz to the signal-antenna receivers.

\begin{figure}[htbp]
\centering
\includegraphics[scale=0.1]{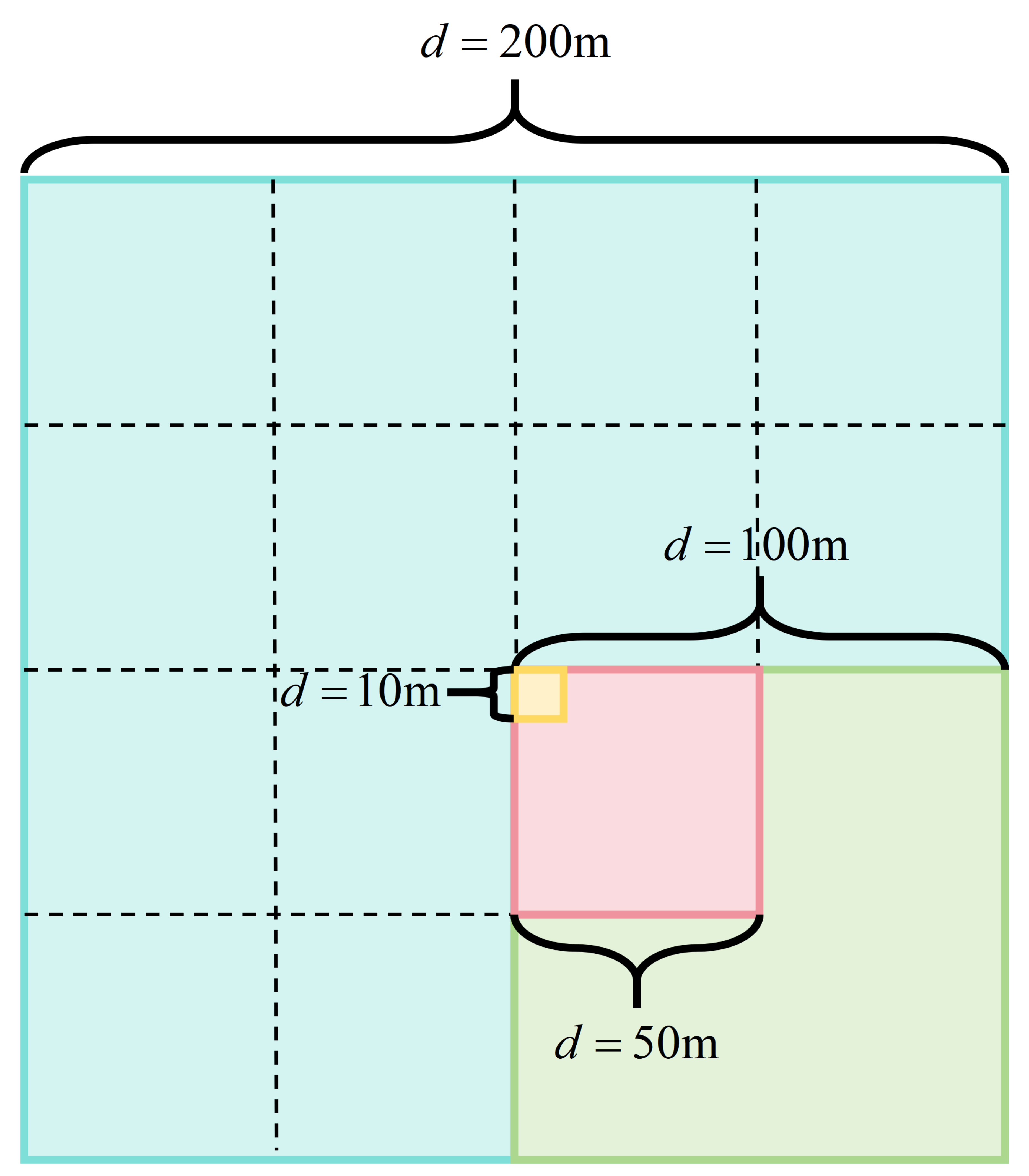}
\caption{Selected subregions with different grid sizes.}
\label{figgridsizes}
\end{figure}
As analyzed in Section IV-B, the proposed CSFM in this paper comprises two key components of information. The first component is the network parameters of the channel denoisers stored in each grid, while the second component consists of historical channel data sampled at discrete points. To evaluate the performance of denoisers with different grid sizes, this paper uniformly partitioned the target area depicted in Fig. \ref{figWIscenario} into grids with dimensions of $d=10$, $50$, $100$, and $200$ meters, respectively. The partitioning result is presented in Fig. \ref{figgridsizes}. It is evident that an increase in grid size leads to a significant augmentation in the amount of channel data within each grid, as well as a greater complexity and diversity of environmental characteristics. To thoroughly balance denoising accuracy, computational complexity, and training time, this paper selectes the yellow, red, green, and blue grids from Fig. \ref{figgridsizes} as the subjects for the 10-meter, 50-meter, 100-meter, and 200-meter grid sizes, respectively. Subsequently, 500 channel samples were randomly extracted from each of the four selected grids to construct the testing dataset. Additionally, 1000, 20000, 40000, and 100000 channel samples are randomly selected from the remaining channel samples of the four grids, respectively, to train the denoiser network parameters. For the historical channel data at discrete sampling points, channel samples were preserved at intervals of 1 meter.

The pilot signals at the transmitter are generated based on DFT matrices. A DFT matrix $\mathbf{F}\in\mathbb{C}^{M\times M}$ is generated according to the number of antennas $M$ at the BS. When the number of pilot less than the number of antenna, i.e., $\tau < M$, $\tau$ rows are uniformly selected from $\mathbf{F}$ to form the pilot signal $\mathbf{X}$. When $\tau = M$, the entire DFT matrix $\mathbf{F}$ is directly employed as the transmitted pilot signal $\mathbf{X}$.

\subsection{Benchmark Algorithms and Performance Metric}
In complex simulation scenarios, the $P_{h|y}\left(\mathbf{h}|\mathbf{y}\right)$ required by the MMSE algorithm is difficult to obtain, and the channel estimation problem of equation \eqref{eq9} can not be solved directly. Therefore, three primary benchmarking schemes are considered in this paper, i.e., LS/ML, LMMSE, and CSFM-NN channel estimation.
\subsubsection{LS/ML}
Due to the number of orthogonal pilots being not greater than the number of antennas at the BS i.e., $\tau \le M$. The pseudo inverse in \eqref{eq5} is $\mathbf{X}^\dagger=\mathbf{X}^{H}\left(\mathbf{X}\mathbf{X}^{H}\right)^{-1}$.
\subsubsection{LMMSE}
This channel estimation algorithm is solved using equation \eqref{eq8}, where the channel mean and covariance matrix are obtained as follows: In the scenario depicted in Fig. \ref{figWIscenario}, the entire area is uniformly divided into grids of size $1$ meter, and first-order and second-order statistical processing is performed on all channels within each grid. Based on \cite{CCM}, the channel statistical data obtained by this way is very close to the groundtruth values.
\subsubsection{CSFM-NN}
This channel estimation involves comparing the location of any channel in the test channel dataset with the discrete channel locations stored in the CSFM, and the channel is estimated using the NN algorithm.

Besides, in this paper, the normalized mean square error (NMSE) is employed as the pivotal performance metric to quantify the accuracy of channel estimation, expressed as follows
\begin{equation}
\text{NMSE}=\frac{1}{K}\sum_{k=1}^{K}\frac{||\mathbf{h}_{k}-\widehat{\mathbf{h}}_{k}||^2}{||\mathbf{h}_{k}||^2},\label{eq27}
\end{equation}
where $K$ is the number of channels in the test dataset. $\mathbf{h}_{k}$ and $\widehat{\mathbf{h}}_{k}$ are the groundtruth channel and the estimated channel of the $k$-th test data, respectively.

\subsection{Numerical Results}
Based on the analyses presented in Section IV-C and VI-A, the parameter configurations for the simulations conducted in this paper are summarized in Table \ref{table2}.
\renewcommand{\arraystretch}{1.4}
\begin{table}[htbp]
\centering
\caption{Summary of the parameter configurations.}
\label{table2}
\begin{tabular}{|c|c|}
\hline
\textbf{Parameters} & \textbf{Value} \\ \hline
Carrier frequency & $28$ GHz\\ \hline
Transmitted pilot power: $\rho$ & $1$\\ \hline
Number of pilot signals: $\tau$ & $16$\\ \hline
Number of antennas at the BS: $M$ & $16 \times 16$\\ \hline
Number of channels in test dataset: $K$ & $500$\\ \hline
Number of channels in training dataset ($d=10$) & $1000$\\ \hline
Number of channels in training dataset ($d=50$) & $20000$\\ \hline
Number of channels in training dataset ($d=100$) & $40000$\\ \hline
Number of channels in training dataset ($d=200$) & $100000$\\ \hline
\end{tabular}
\end{table}

\begin{figure}[htbp]
\centering
\includegraphics[scale=0.52]{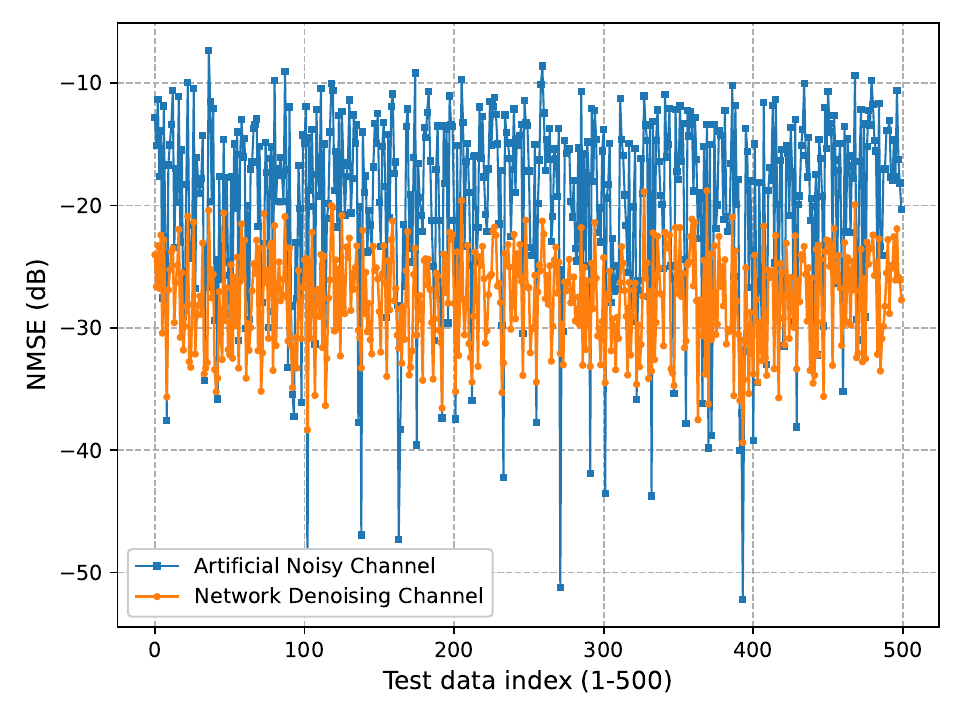}
\caption{Performance improvement of the designed denoiser ($d=50$ meters).}
\label{figNMSEDenoiserVSaddnoise}
\end{figure}

Fig. \ref{figNMSEDenoiserVSaddnoise} illustrates the performance of the denoiser within the selected red grid with the grid size $d=50$ meters, following 25000 epochs. The horizontal axis represents the index numbers of channels in the test set, while the vertical axis denotes the NMSE, which is computed according to equation \eqref{eq27}. The curve labeled ``Artificial Noisy Channel'' represents the NMSE between the original noise-free channel generated by ray-tracing, and the noisy channel generated by adding Gaussian noise with random variance to the noise-free channel. The curve labeled ``Network Denoising Channel'' represents the NMSE between the original noise-free channel and the channel obtained after a single-step processing of the noisy channel through the trained denoiser. The designed denoiser achieves an average performance improvement of $7.84$ dB. Under the same number of training epochs, the single-step average denoising performance improvements achieved by the denoisers within the grids corresponding to $d=100$ and $200$ meters are $7.11$ dB and $7.01$ dB, respectively. Specifically, due to the limited channel data within the grid of $d=10$ meters, the channel denoiser trained for the same number of epochs fails to achieve effective channel denoising. The NMSE of the single-step denoised channel is, on average, 5.17 dB higher than that of the artificially noisy channel.

\begin{figure}[htbp]
\centering
\includegraphics[scale=0.54]{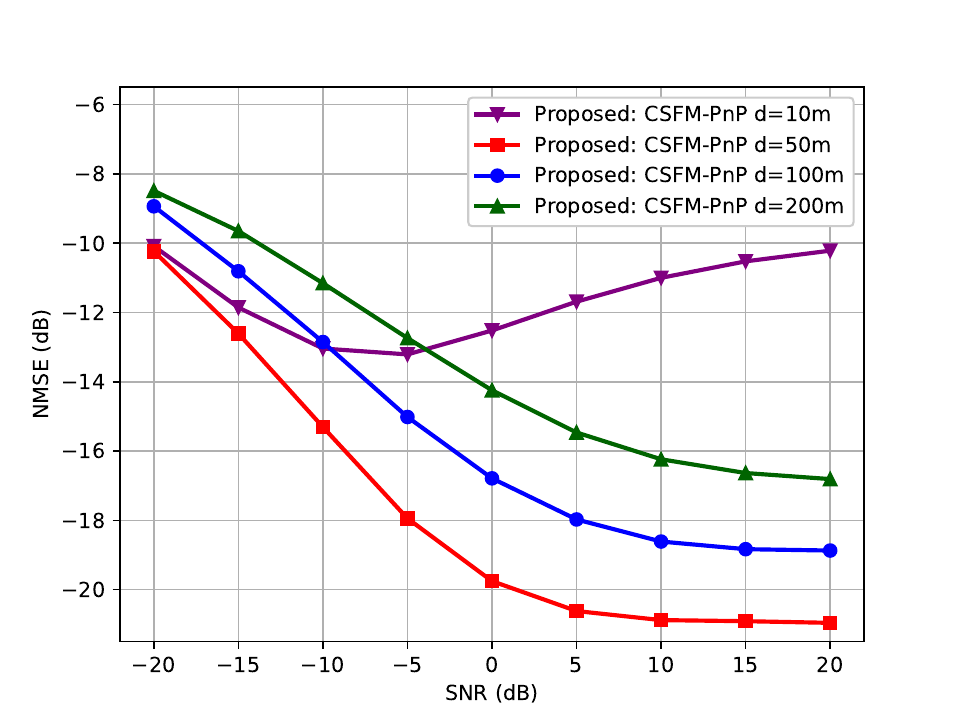}
\caption{Performance comparison of denoisers with varying grid sizes on the proposed channel estimation algorithm ($\tau=16$).}
\label{figNMSEfourgridsizes}
\end{figure}

Fig. \ref{figNMSEfourgridsizes} illustrates the performance comparison of denoisers with varying grid sizes on the proposed CSFM-PnP-based environment-aware channel estimation algorithm. In Algorithm \ref{alg1}, the values of parameters $\alpha$, $\alpha^{\prime}$, $\beta$, and $I$ are set to $\frac{1}{2\left(1+e^{-log_{10}\left(SNR\right)}\right)}$, $0.1875$, $0.0001$, and $10$, respectively. The horizontal axis represents the SNR of the measured pilot signal power to the noise power in equation \eqref{eq3}, while the vertical axis denotes the NMSE of the channel estimation in equation \eqref{eq27}. The results indicate that reducing the size of the subregions indeed enhances the performance of channel estimation. However, when the subregions become excessively small, the performance of channel estimation deteriorates. This is attributed to the fact that the limited amount of historical channel data within small grid may be insufficient to effectively train deep learning models, thereby degrading model performance. Specifically, when the denoiser trained within a grid of $d=10$ meters is directly applied to the proposed CSFM-PnP channel estimation algorithm, its performance initially increases and then decreases with the rise of the SNR. This behavior is primarily attributed to the parameter selection in this paper. As the SNR increases, the weight of the channel denoiser in \eqref{eq24} also increases. Since the denoiser within the $d=10$ meter grid is unable to perform effective denoising and instead exhibits a noise-enhancing characteristic, the observed trend in Fig. \ref{figNMSEfourgridsizes} emerges.

\begin{figure}[htbp]
\centering
\includegraphics[scale=0.54]{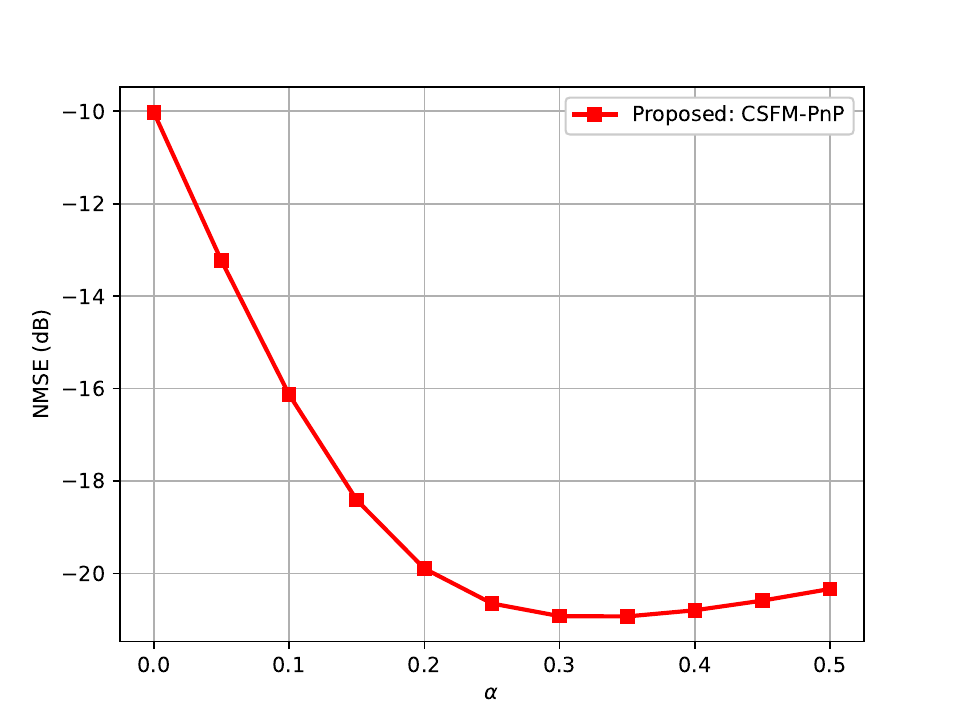}
\caption{Impact of SD algorithm step-size coefficient $\alpha$ on the proposed channel estimation algroithm in no pilot signal scenario ($d=50$ meters).}
\label{figNMSEwithoutpilot}
\end{figure}
In the challenging scenario of channel generation without any pilot signals ($\mathbf{X} =\mathbf{0}$), classical LS and ML methods are ineffective. Equation \eqref{eq8} shows that the LMMSE method yields $\widehat{\mathbf{h}}_{\text{LMMSE}}=\overline{\mathbf{h}}$, with an NMSE of $0.141$ dB, independent of SNR. Similarly, CSFM-NN achieves an NMSE of $-9.987$ dB, also SNR-independent. Besides, equations \eqref{eq18} and \eqref{eq24} reveal that when $\mathbf{X} =\mathbf{0}$, the proposed CSFM-PnP algorithm operates as a generative channel estimation method. Its performance is determined by the step size coefficient $\alpha$ of the SD algorithm and is independent of the received SNR. Therefore, Fig. \ref{figNMSEwithoutpilot} illustrates the impact of the SD algorithm step size coefficient $\alpha$ on the proposed channel estimation algorithm in the no pilot signal scenario. The denoiser used in the CSFM-PnP algorithm is derived from the selected region with grid size $d=50$ meters, with parameters $\alpha^{\prime}$, $\beta$, and $I$ in Algorithm \ref{alg1} set to 0.1875, 0.0001, and 10, respectively. It is evident that the best channel estimation result is achieved when $\alpha\approx 0.3$. And for the challenging scenario without pilot signals, the proposed algorithm significantly outperforms traditional baseline methods.

\begin{figure}[htbp]
\centering
\includegraphics[scale=0.54]{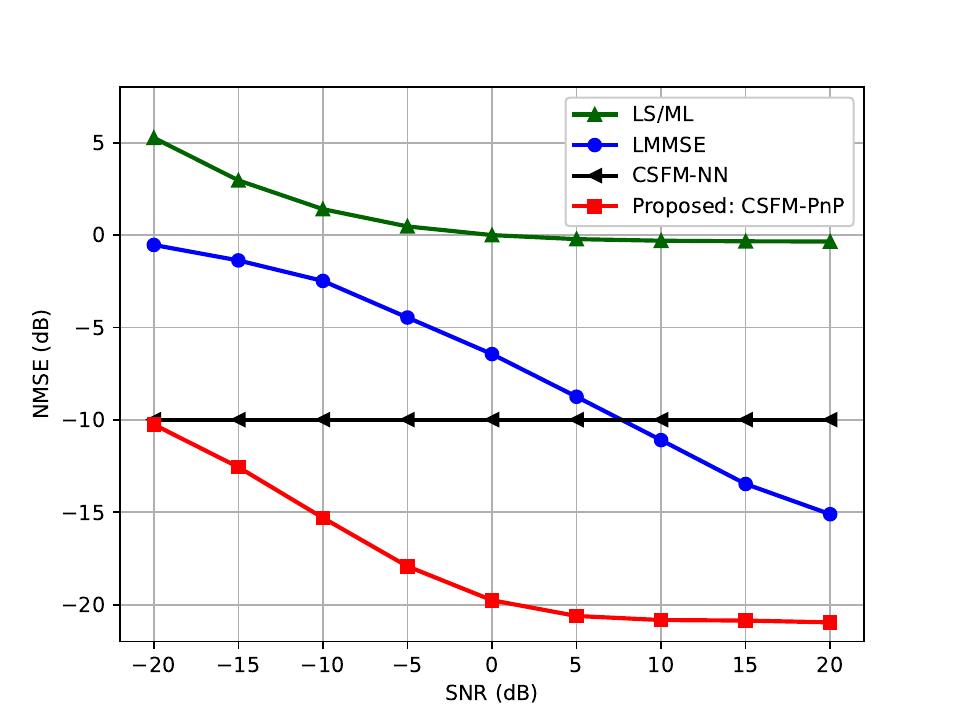}
\caption{NMSE comparison of different channel estimation algorithms ($d=50$ meters and $\tau=16$).}
\label{figNMSE16pilot50mgrid}
\end{figure}

Fig. \ref{figNMSE16pilot50mgrid} illustrates the channel estimation results for the proposed CSFM-PnP channel estimation algorithm and other benchmark algorithms, where the number of pilot signals is 16 and the denoiser originates from the selected red grid in Fig. \ref{figgridsizes} with grid size $d=50$ meters. It can be inferred that the proposed CSFM-PnP based environment-aware channel estimation algorithm significantly outperforms the three benchmark algorithms. The LS/ML channel estimation algorithm perpetually underperforms, primarily attributable to the scarcity of pilot signals and the deleterious effects of measurement noise. The CSFM-NN algorithm, which directly retrieves channel information from the CSFM using the NN algorithm, is unaffected by the SNR of the observations. While LMMSE improves with SNR, it never surpasses the proposed method. This is because real-world channel distributions often deviate from the Gaussian assumption. In contrast, the denoiser within the proposed algorithm effectively handles the complex channel prior distribution, thereby achieving higher-quality channel estimation results.

\begin{figure}[htbp]
\centering
\includegraphics[scale=0.54]{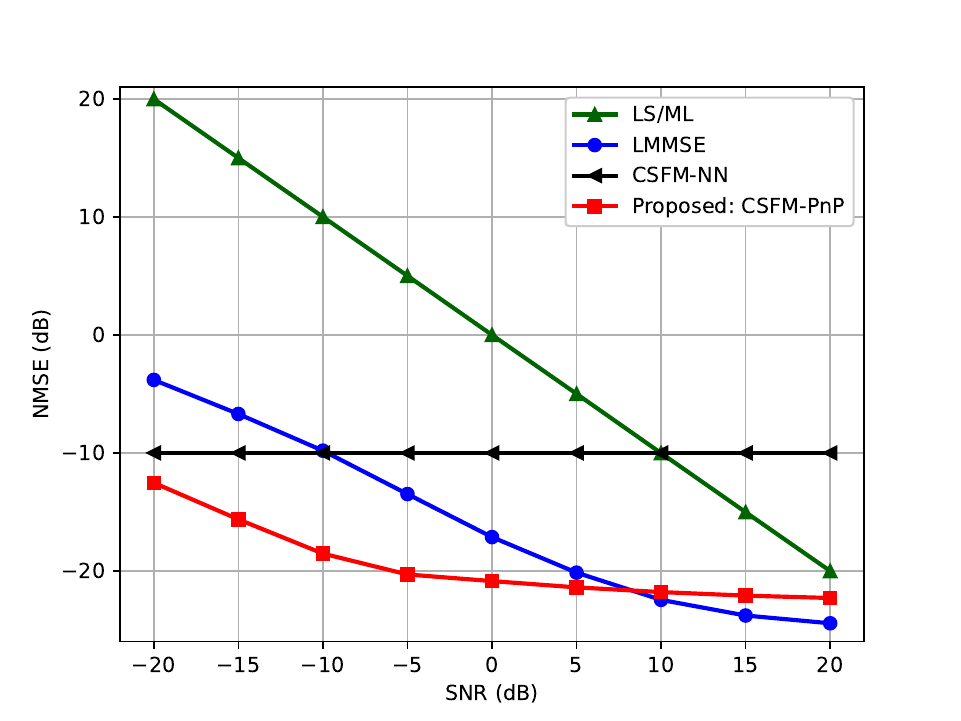}
\caption{NMSE comparison of different channel estimation algorithms ($d=50$ meters and $\tau=256$).}
\label{figNMSE256pilot50mgrid}
\end{figure}

Distinguished from Fig. \ref{figNMSE16pilot50mgrid}, Fig. \ref{figNMSE256pilot50mgrid} offers an analysis of the performance of several channel estimation algorithms in the context of abundant pilot sequences. The parameters $\alpha$, $\alpha^{\prime}$, $\beta$, and $I$ used in the proposed CSFM-PnP channel estimation algorithm are set to $\frac{2}{5\left(1+e^{-log_{10}\left(SNR\right)}\right)}$, $2.25$, $0.0001$, and $20$, respectively. As depicted in Fig. \ref{figNMSE256pilot50mgrid}, the proposed algorithm attains satisfactory channel estimation outcomes in low SNR scenarios, while consistently outperforming conventional baseline methods. Furthermore, by comparing Fig. \ref{figNMSE16pilot50mgrid} and Fig. \ref{figNMSE256pilot50mgrid}, it can be observed that increasing the number of pilot observation signals effectively enhances the accuracy of channel estimation using the proposed CSFM-PnP algorithm.

\section{Conclusion}
In this paper, we studied low-overhead and high-precision large-dimensional channel estimation in the four aforementioned challenging 6G scenarios. Based on the novel concept of CKM, we propose a novel CSFM-PnP based environment-aware channel estimation algorithm within a regularized MAP framework. Initially, we decouple and simplify the original optimization problem using the PnP algorithm. Subsequently, we employ Tweedie's formula to connect the channel score function with the channel denoiser, effectively handling the difficult channel prior distribution term. The channel score function is then approximated using an environment-aware channel denoiser provided by the CSFM, achieving efficient channel estimation. Besides,this paper also details the CSFM construction process. Finally, numerical results demonstrate that our algorithm significantly outperforms existing classical baseline schemes in these challenging scenarios.

\bibliographystyle{IEEEtran}
\bibliography{reference.bib}

\end{document}